\documentclass[draftclsnofoot, onecolumn, comsoc, 12pt]{IEEEtran} 
\usepackage{upgreek}
\usepackage[caption=false,font=normalsize,labelfont=sf,textfont=sf]{subfig}
\usepackage{array}
\usepackage{textcomp}
\usepackage{stfloats}
\usepackage{url}
\usepackage{verbatim}
\usepackage{colortbl}
\usepackage{diagbox}
\usepackage{adjustbox}
\usepackage{float}
\usepackage{lipsum}
\usepackage{boldline}
\usepackage{hhline}
\usepackage[dvipsnames]{xcolor}

\usepackage{graphicx,epsfig}
\usepackage[noadjust]{cite}
\usepackage{mcite}
\usepackage{amsfonts,helvet}
\usepackage{fancyhdr}
\usepackage{threeparttable}
\usepackage{epsf,epsfig}
\usepackage{amsthm}
\usepackage{amsmath}
\usepackage{siunitx}
\usepackage{amssymb}
\usepackage{dsfont}
\usepackage{color}
\usepackage[linesnumbered,ruled]{algorithm2e}
\usepackage{algpseudocode}
\usepackage{enumerate}
\usepackage{cancel}
\usepackage{comment}
\usepackage{booktabs}
\usepackage{multirow}

\newtheorem{case}{Case}

\newtheorem{remark}{Remark}

\SetKwInput{KwInput}{Input}
\SetKwInput{KwInt}{Initialize}
\SetKwInput{KwOutput}{Output}
\SetKwProg{Fn}{Stage 1}{}{}
\SetKwProg{Fn}{Stage 2}{}{}
\SetKw{Continue}{continue}

\usepackage{eucal}
\usepackage{soul}

\def\BibTeX{{\rm B\kern-.05em{\sc i\kern-.025em b}\kern-.08em
    T\kern-.1667em\lower.7ex\hbox{E}\kern-.125emX}}
\usepackage{balance}
\begin{document}
\title{{A Double-Difference Doppler Shift-Based Positioning Framework with Ephemeris Error Correction of LEO Satellites
}}
\author{Md. Ali Hasan, M. Humayun Kabir, Md. Shafiqul Islam, \\Sangmin Han, and Wonjae Shin
%
%
%
%
\thanks{Md. Ali Hasan, Sangmin Han, and Wonjae Shin are with the School of Electrical Engineering, Korea University, Seoul 02841, Republic of Korea (e-mail: alihasan@korea.ac.kr; smhan22@korea.ac.kr; wjshin@korea.ac.kr). M. Humayun Kabir is with the Department of Electrical and
Electronic Engineering, Islamic University, Kushtia 7003, Bangladesh (e-mail:
\mbox{humayun@eee.iu.ac.bd}). Md. Shafiqul Islam is with the Department of Computer Science and Engineering, Bangladesh University of Business and Technology, Dhaka, Bangladesh (e-mail:
\mbox{msislam@bubt.edu.bd}).}
}


\maketitle\vspace{-10mm}
\begin{abstract}
In signals of opportunity (SOPs)-based positioning utilizing low Earth orbit (LEO) satellites, ephemeris data derived from two-line element files can introduce increasing error over time. To handle the erroneous measurement, an additional base receiver with a known position is often used to compensate for the effect of ephemeris error when positioning the user terminal (UT). However, this approach is insufficient for the long baseline (the distance between the base receiver and UT) as it fails to adequately correct Doppler shift measurement errors caused by ephemeris inaccuracies, resulting in degraded positioning performance. Moreover, the lack of clock synchronization between the base receiver and UT exacerbates erroneous Doppler shift measurements. To address these challenges, we put forth a robust double-difference Doppler shift-based positioning framework, coined \textit{3DPose}, to handle the clock synchronization issue between the base receiver and UT, and positioning degradation due to the long baseline. The proposed 3DPose framework leverages double-difference Doppler shift measurements to eliminate the clock synchronization issue and incorporates a novel ephemeris error correction algorithm to enhance UT positioning accuracy in case of the long baseline. The algorithm specifically characterizes and corrects the Doppler shift measurement errors arising from erroneous ephemeris data, focusing on satellite position errors in the tangential direction.
To validate the effectiveness of the proposed framework, we conduct comparative analyses across three different scenarios, contrasting its performance with the existing differential Doppler positioning method. The results demonstrate that the proposed 3DPose framework achieves an average reduction of $90\%$ in 3-dimensional positioning errors compared to the existing differential Doppler approach.
\end{abstract}


\section{Introduction}
\label{sec: Introduction}
\IEEEPARstart{W}{ith} 
the proliferation of Internet of Things (IoT) devices and their integration into various sectors, including transportation, industries, communication, and smart cities, the demand for reliable outdoor positioning systems is becoming even more pronounced. 
As the global navigation satellite system (GNSS) is the dominating outdoor positioning system nowadays, much of the past research has been primarily devoted to GNSS for outdoor positioning in IoT applications \cite{10302358,9844131,10040628,9945837,chen2013localization,7302503}.
{Typically, achieving a position in the global reference frame involves integrating an additional GNSS chipset into IoT devices \cite{9170636}.}
{Observable is the measurement of the quantities of interest that are used in position estimation, excluding errors \cite{noureldin2012fundamentals}. 
GNSS utilizes two main kinds of observables, pseudorange and Doppler shift, to estimate UT position. 
Pseudorange is calculated using either the code phase technique \cite{ward1981inside} or the carrier phase technique \cite{noureldin2012fundamentals}. 
The signal propagation time is estimated using the code phase technique and converted to pseudorange by multiplying signal velocity with the estimated signal propagation time. Meanwhile, in the carrier phase technique, the fractional phase of a carrier signal is estimated, and the unknown full cycle of the carrier signal is referred to as integer ambiguity. After resolving integer ambiguity, pseudorange is obtained by multiplying carrier wavelength with the carrier phase measurement.}
The positioning accuracy of the pseudorange-based GNSS positioning depends on the signal quality and satellites-to-user geometry \cite{kabir2022performance}.

The high altitude and lower dynamics of geostationary Earth orbit (GEO) and medium Earth orbit (MEO) satellites cause lower signal strength and less Doppler shift{, which can not provide good positioning accuracy \cite{jiang2022leo}}. 
{GNSS signal suffers from jamming and spoofing while broadcasting navigation messages due to the lower signal strength \cite{Spensnavi.537}. 
In the presence of jamming, the signal can not be demodulated to get the navigation message. Due to spoofing signals, the positioning accuracy of the UT {is} degraded.}
The GNSS signal may be blocked in some geographical areas, such as large urban canyons, deserts, and forests.
It has poor geometry between the satellite and UT, with a smaller number of satellites in view.
It also suffers from multipath and non-line-of-sight (NLOS) signals in an urban canyon environment. In addition, the pseudorange yields large measurement noise in the multipath and NLOS environment \cite{5751245,van1992multipath}.
In contrast, the lower orbital altitude and higher dynamics of low Earth orbit (LEO) satellites cause considerable signal strength ($\num{30}$ dB larger than the MEO satellite) and large Doppler shift opportunity. 
Doppler shift can be measured without demodulating the carrier signal, and it is less affected by multipath and NLOS environment \cite{9262624,gowdayyanadoddi2015ray}.
Moreover, sufficient LEO satellite constellations ensure suitable satellite-to-user geometry that provides better positioning accuracy. 
Therefore, Doppler shift-based positioning using LEO satellites could be an alternative to the pseudorange-based conventional GNSS in GNSS-challenged environments.

The early transit satellite system can calculate UT position by measuring Doppler shift using a single satellite at $\num{1000}$ km altitude. 
It needs almost $\num{12}$-$\num{14}$ minutes to estimate the UT position with an accuracy of $\num{200}$-$\num{300}$ meter by performing Doppler curve fitting, which is not suitable for single point positioning \cite{parkinson1996introduction}.

Nowadays, many satellites are deployed in the LEO satellite constellation
{by companies such as SpaceX, OneWeb, and Amazon \cite{LI201973}}, which results in multiple satellites being visible to the receiver. 
UT can use Doppler shift measurement from multiple LEO satellites to estimate its position with an appropriate algorithm.
At present, satellites from the LEO constellation are mainly used to provide broadband internet services and do not broadcast any navigation messages for positioning.
To overcome this deficiency, several works focus on space-based signals of opportunity (SOPs) that utilize radio signals of LEO satellites to measure Doppler shift to estimate UT position \cite{BENZERROUK2019496, shi2023revisiting, 6851486, jiang2022leo, 8682554,10415436,5507270,farhangian2021opportunistic,dif9843493}.
{Utilizing an opportunistic approach, the UT only uses the downlink LEO signals, which maintain the privacy of the user \cite{10009900}.}

{A novel coarse positioning system is introduced in \cite{10415436} that uses an angle-of-arrival (AoA) estimate of SOPs from a spacecraft in low Earth orbit. 
The AoA estimation is done in real-time using phase interferometry through a dedicated hardware architecture.
A geometric simplified model that connects the AoA estimate from the reference and target nodes to the planar cartesian coordinate is used to estimate the target's position with respect to the reference node. The system achieves an error in the order of units of kilometers.}
{Benzerrouk \textit{et al.} \cite{BENZERROUK2019496} use Iridium Next LEO satellites to develop a search and rescue (SAR) positioning information system based on SOPs. They utilize nonlinear least squares and filtering algorithms such as extended Kalman filter (EKF) for positioning solutions.
Another work \cite{8682554} focuses on EKF in Doppler shift-based positioning and reported a position accuracy of 11 m in a simulation environment using $\num{25}$ LEO satellites.
In addition, an experimental environment is set up using two Orbcomm satellites with a reported positioning accuracy of 360 m.}

{To improve the positioning performance in GNSS denied environment, the LEO satellite system is coupled with GNSS.
The assistance of Doppler shift measurement from the LEO satellite with GNSS pseudorange measurement is proposed in \cite{6851486}. They use a single IRIDIUM and three global positioning system (GPS) satellites to get the position of a stationary UT utilizing the TLE file to predict the orbital parameters of the LEO satellite.
In addition, the signals from the LEO satellite in SOPs mode are used to assist the GNSS system where the GNSS signal is blocked due to jamming by interference sources \cite{jiang2022leo}.
They utilize Doppler shift measurement from the LEO satellites with pseudorange and Doppler shift obtained from GNSS. They also analyze the sensitivity of position accuracy with the measurement error and show that positioning accuracy degraded due to the measurement error of the GNSS Doppler shift. 
During GNSS outages, a simultaneous tracking and navigation (STAN) framework is introduced in \cite{kassas2024ad} utilizing LEO satellites.
Moreover, Shi \textit{et al.} \cite{shi2023revisiting} propose a Doppler-only point-solution algorithm using LEO satellites. They estimate the receiver position, velocity, and clock drift and analyze the effect of atmospheric error on positioning accuracy. 
They claim that the position accuracy degrades about dozens of meters without atmospheric correction.}

{Unlike the GNSS system, LEO satellites do not necessarily broadcast their orbital parameters, including their position and velocity, as they are not designed for navigation purposes \cite{kassas2024ad}.} 
In this case, the publicly available two-line element (TLE) files {provided by the North American Aerospace Defense Command (NORAD)} and orbit propagator models, like simplified general perturbations 4 (SGP4), are used to estimate satellite ephemeris data. However, the ephemeris data obtained by TLE files and the SGP4 orbit propagator model are not always perfect. Some recent works \cite{morales2019orbit} and \cite{10288195} report satellite position error and velocity error around $\num{3}$ km and $\num{3}$ m/s, respectively.
These position and velocity errors induce Doppler shift measurement errors, which are known as Doppler shift errors due to ephemeris errors. 
The effect of this ephemeris error comes up with a significant positioning error of the UT that can be more than several km \cite{shi2023revisiting}.

To compensate for the Doppler shift measurement error due to the ephemeris error, Farhangian \textit{et al.} \cite{farhangian2021opportunistic} couple inertial navigation system (INS) with Doppler shift measurements. 
{This method improves the positioning accuracy up to $\num{180}$ \% compared with the Doppler shift-based positioning method without INS.
A tightly coupled GNSS/INS/LiDAR integration method is proposed in \cite{9844131} to support the emerging IoT application (e.g., self-driving car).}
However, navigation information obtained from INS accumulates errors over time \cite{9955423}.
An additional base receiver with a known position is introduced in \cite{dif9843493} together with UT to collect Doppler shift measurements.
This method minimizes the Doppler shift measurement error due to ephemeris error as well as eliminates satellite clock synchronization issues.
Introducing a base receiver with UT increases costs as it needs installation and maintenance. However, for large-scale deployment, it is convenient to sacrifice the cost of the base receiver for accuracy.
Since all the clocks used in LEO satellites, base receiver, and UT are not synchronized, the Doppler shift measurement error related to clock drift needs to be considered \cite{8682554}.
{The} difference {between} Doppler shift measurements {of} the base receiver and UT is considered as a single-difference Doppler shift measurement.
By utilizing single-difference Doppler shift measurement, the UT can reduce the effect of the ephemeris error and eliminate the error related to satellite clock drift.

\subsection{Motivations and Contributions}
Despite this positive approach \cite{dif9843493}, the effect of the clock synchronization issue between the base receiver and UT still remains and needs to be addressed.
Moreover, The measurement error of Doppler shift for the ephemeris error is non-linear with the distance between the base receiver and UT.
The distance between the base receiver position and the UT position is considered a baseline.
{By increasing the baseline, 
{the Doppler shift measurement errors due to the ephemeris error are not always equal for the base receiver and UT.
Thus,} the uncertainty of ephemeris error is not minimized in differential Doppler positioning \cite{dif9843493}, which causes inadequate positioning accuracy.} 
{It is necessary to design a system that can address this uncertainty to provide robust positioning.}
To tackle these challenges, we propose a precise and robust double-difference Doppler shift-based positioning framework, named \textit{3DPose}, to handle the clock synchronization issue of the base receiver and UT as well as the positioning degradation due to the long baseline.
The difference of single-difference Doppler shift measurement between the satellites is considered a double-difference Doppler shift measurement. 
The main contributions of this article are summarized as follows.
\begin{enumerate}
\item We propose a 3DPose framework using LEO satellites that utilizes double-difference Doppler shift measurements to eliminate the clock synchronization effect between the base receiver and UT.
\item We put forth a novel ephemeris error correction algorithm that characterizes and corrects the Doppler shift measurement error due to the ephemeris error at the estimated UT position to reduce the positioning error in the case of the long baseline. 
The estimated UT position is obtained from the double-difference Doppler shift-based positioning framework. To handle the non-linear Doppler shift measurement error, we reformulate the Doppler shift equation considering LEO satellite position error acting along the tangential direction utilizing Taylor approximation.
{Therefore, the 3DPose framework provides robustness in terms of Doppler shift measurement error.}

\item Unlike the existing Doppler shift-based positioning methods focusing on stationary UT, we drive a Jacobian matrix including the position and velocity components of a moving UT in our proposed 3DPose framework.

\item We validate the proposed 3DPose framework and compare it with the existing differential Doppler positioning method \cite{dif9843493}.
Our proposed 3DPose framework achieves higher positioning accuracy in all scenarios and specifically improves 3-dimensional positioning accuracy by an average of ${90}$ \% than the existing method \cite{dif9843493}.
{The superior positioning accuracy of the proposed 3DPose framework is due to the novel ephemeris error correction algorithm that minimizes the effect of ephemeris error in UT positioning.}

\end{enumerate}

\subsection{Organization and Notations}

The remainder of this study is organized as follows. Section II discusses the details of the system model. Section {III} presents the robust double-difference Doppler shift-based positioning framework. Section {IV} describes the results and discussions. Finally, Section {V} concludes the study and discusses future work.

\textit{Notations:} Bold lowercase and uppercase letters represent vectors and matrices, respectively. $\|\cdot\|$ denotes L2-norm, and $[\cdot]^T$ represent the transpose of a vector (matrix).

\section{System Model}
\label{sec: System_Model}
\begin{figure}[t]
    \centering
    \includegraphics[width=0.9\linewidth]{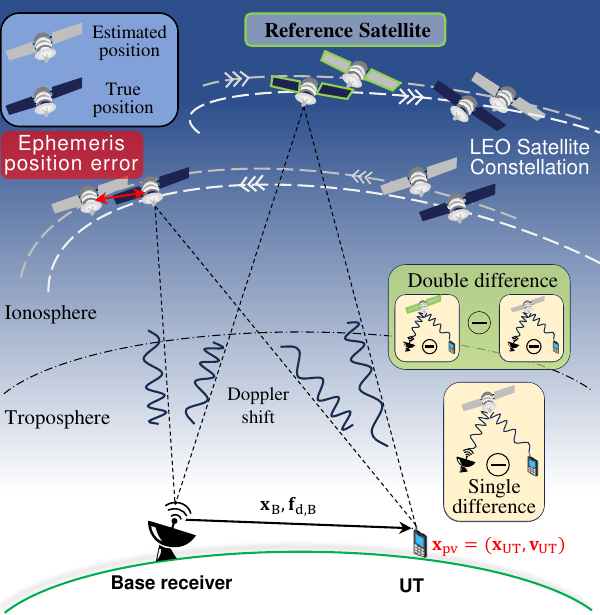}
    \caption{System model of the proposed 3DPose framework.}
    \label{fig: System_model}
\end{figure}

We consider a stationary base receiver with a known position ($\mathbf{x}_{\sf B}$) and a moving UT with an unknown position ($\mathbf{x}_{\sf UT}$) and velocity ($\mathbf{v}_{\sf UT}$). The base receiver and UT simultaneously measure the Doppler shift received from multiple common LEO satellites in view, as shown in Fig.~\ref{fig: System_model}. 
The base receiver is capable of sending the Doppler shift measurements ($\mathbf{f}_{\sf d, B}$) along with its known position to the UT using any kind of communication technology. We assume that there is no communication delay between the base receiver and UT. 
{However, in real-world applications, there is a communication delay between the base receiver and UT based on the type of communication technology. For 5G networks, the communication delay is less than a millisecond \cite{10011409}, which can cause a millimeter-level positioning error of UT. This type of communication delay can not create a significant impact on the real-world applicability of the proposed 3DPose framework.}
We consider the $L$ number of visible satellites for both the base receiver and UT sides at a specific time. The true position and velocity of the visible satellite are denoted as $\mathbf{x}^\ell_{\sf sat}$ and $\mathbf{v}^\ell_{\sf sat}$ in the Earth-centered Earth-fixed (ECEF) coordinate frame, where $\ell$ is the index of the satellite. 

Doppler frequency depends on the relative velocity between the UT and $\ell$-th satellite in the line-of-sight (LOS) direction. The Doppler shift $f_{{\sf d},{\sf UT}}^{\ell}$ is defined as 
\begin{equation}
    f_{{\sf d},{\sf UT}}^{\ell} =f_{\sf R,UT}^\ell-f_{\sf T}^\ell=\pm\frac{\|\mathbf{v}^\ell_{\sf LOS}\|}{{\sf c}}f_{\sf T}^\ell,
 \label{eq: 1}
\end{equation}
where $f_{\sf R,UT}^\ell$, $f_{\sf T}^\ell$, $\mathbf{v}^\ell_{\sf LOS}$, and ${\sf c}$ are the receive frequency, transmit frequency, relative velocity in the LOS direction, and speed of radio frequency (RF) signal, respectively. The Doppler shift is positive if the LEO satellite and UT are moving toward, while negative if they are moving away.

Rearranging \eqref{eq: 1}, we get
\begin{equation}
\pm f_{{\sf d},{\sf UT}}^{\ell}\lambda_{\sf T}^\ell= \|\mathbf{v}^\ell_{\sf LOS}\|,
\label{eq: 2}
\end{equation}
where $\lambda_{\sf T}^\ell={\sf c}/f_{\sf T}^\ell$ is the wavelength of the transmitted signal from $\ell$-th LEO satellite. The measured LOS velocity on the UT side contains errors due to the receiver clock drift, satellite clock drift, atmospheric delay rate, and other measurement errors. Thus, the multiplication of Doppler shift and transmitted signal wavelength is defined as pseudorange rate ($\Dot{\rho}_{\sf UT}^\ell$) between UT and $\ell$-th satellite \cite{9110186,6851486}, which yields
\begin{equation}
   -f_{{\sf d, UT}}^{\ell}\lambda^\ell_{\sf T}=\Dot{\rho}_{\sf UT}^\ell.
\label{eq: 3}
\end{equation}

The pseudorange ($\rho_{\sf UT}^\ell$) is defined as the range between \mbox{$\ell$-th} satellite and UT, which is measured by the receiver before the determination of time bias and delay corrections \cite{ward1981inside}. Thus, the pseudorange is as follows.
\begin{align}
    \rho^\ell_{\sf UT} &= \|\mathbf{x}_{\sf sat}^{\ell}-\mathbf{x}_{{\sf UT}}\|+ 
    {\sf c}\big(a_{0,{\sf UT}}+a_{1,{\sf UT}}\left(t_{{\sf R},{\sf UT}}-t_{0,{\sf UT}}\right) \notag \\
    &+\frac{1}{2}a_{2,{\sf UT}}\left(t_{{\sf R},{\sf UT}}-t_{0,{\sf UT}}\right)^2+\psi_{\sf UT}\left(t_{{\sf R, UT}}\right)\big)
    - {\sf c}\big(a^\ell_0 \notag \\
    &+a^\ell_1\left(t^\ell_{\sf T}-t^\ell_0\right)+\frac{1}{2}a^\ell_2 
    \left(t^\ell_{\sf T}-t^\ell_0\right)^2+\psi^\ell\left(t^\ell_{\sf T}\right)\big) 
    +\Delta T^\ell_{\sf UT} \notag \\
    &+\Delta I^\ell_{\sf UT}+M^{\ell}_{\sf UT}+\epsilon_{\rho^\ell_{\sf UT}},
    \label{eq: 4}
\end{align}
where $a_{0,{\sf UT}}$, $a_{1,{\sf UT}}$, and $a_{2,{\sf UT}}$ are the clock bias, clock drift, and frequency drift of UT at initial time $t_{0,{\sf UT}}$, respectively. $a^\ell_0$, $a^\ell_1$, and $a^\ell_2$ are the clock bias, clock drift, and frequency drift of $\ell$-th satellite at initial time $t^\ell_0$, respectively. $\psi_{\sf UT}\left(t_{\sf R,UT}\right)$ and $\psi^\ell(t^\ell_T)$ are the stochastic noise for UT and the $\ell$-th satellite, respectively. $\Delta T^\ell_{\sf UT} = T^{\ell}_{\sf UT}-\hat{T}^{\ell}_{\sf UT}$, $\Delta I^\ell_{\sf UT} = I^{\ell}_{\sf UT}-\hat{I}^{\ell}_{\sf UT}$, $M^{\ell}_{\sf UT}$, and $\epsilon _{\rho^\ell_{{\sf UT}}}$ are the tropospheric delay error, ionospheric delay error, multipath effect, and measurement noise, respectively.

According to the Saastamoinen model \cite{saastamoinen1972atmospheric}, the estimated tropospheric delay error is obtained as
\begin{equation}
    \hat{T}^\ell_{\sf UT}= \frac{0.002277}{\cos(z)}\times \left[p+(1255/T+0.05)e-1.16 \tan^2(z)\right],
    \label{eq: 5Saastamoinen}
\end{equation}
where $z$, $p$, $e$, and $T$ are the true zenith distance, atmospheric pressure, partial pressure of water vapor, and absolute temperature, respectively.
The estimated Ionospheric delay error is obtained according to the Klobuchar model \cite{klabuchar} as follows.
\begin{align}
    \hat{I}^\ell_{\sf UT} = {\sf c}\times F\times\left[5\times10^{-9}+\sum_{n=0}^{3}\alpha_n \Phi_m^n  \times  \left(1-\frac{a^2}{2}+\frac{a^4}{24}\right)\right],
    \label{eq: 6Klobuchar}
\end{align}
where $a=\frac{2\pi\left(t-50400\right)}{\sum_{n=0}^{3}\beta_n\Phi_m^n}$; $\alpha_n$ and $\beta_n$ are the algorithm parameter. $\Phi_m$ and $t$ are the geomagnetic latitude and local time (s).

The pseudorange rate is the derivative of pseudorange with respect to time and is represented as
\begin{align}
    \Dot{\rho}^\ell_{\sf UT} &= \left(\mathbf{v}_{\sf sat}^{\ell}-\mathbf{v}_{\sf UT}\right)\cdot\frac{\mathbf{x}_{\sf sat}^{\ell}-\mathbf{x}_{\sf UT}}{\|\mathbf{x}_{\sf sat}^{\ell}-\mathbf{x}_{\sf UT}\|}
    +{\sf c}\big(a_{1,{\sf UT}}+a_{2,{\sf UT}}\left(t_{{\sf R,UT}} \right. \notag \\
    &\left.-t_{0,{\sf UT}}\right)
    +\dot\psi_{\sf UT}(t_{{\sf R,UT}})\big)-{\sf c}\big(a^\ell_1+a^\ell_2 
    \left(t^\ell_{\sf T}-t^\ell_0\right) \notag \\
    &+\dot\psi^\ell(t^\ell_{\sf T})\big)
    +\Delta\Dot{T}^{\ell}_{\sf UT}+\Delta\Dot{I}^{\ell}_{\sf UT}+\epsilon _{\Dot{\rho}^\ell_{i}},
    \label{eq: 7}
\end{align} 
where $\Dot{\rho}^{\ell}_{\sf UT}$, $\Delta\Dot{T}^{\ell}_{\sf UT} = \Dot{T}^{\ell}_{\sf UT}-\hat{\Dot{T}}^{\ell}_{\sf UT}$, $\Delta\Dot{I}^{\ell}_{\sf UT} = \Dot{I}^{\ell}_{\sf UT}-\hat{\Dot{I}}^{\ell}_{\sf UT}$, and $\epsilon _{\Dot{\rho}^\ell_{\sf UT}}$ are the pseudorange rate, tropospheric delay rate error, ionospheric delay rate error, and measurement noise, respectively. 
{The delay rate error through the atmosphere (tropospheric/ionospheric) is the rate of change of signal propagation delay times the speed of the RF signal.}

Since the LEO satellites do not necessarily broadcast ephemeris information like GNSS satellites,
the orbit determination algorithm, like the SGP4 model, estimates the orbital parameters of the satellites using the TLE files \cite{vallado2008sgp4}.
Due to the induced error in the TLE file over time (atmospheric drag, solar radiation pressure, third body effect, etc.), the estimated position ($\hat{\mathbf{x}}^\ell_{\sf sat}$) and velocity ($\hat{\mathbf{v}}^\ell_{\sf sat}$) of the satellite differs significantly irrespective from the true position and velocity that referred to as ephemeris error.
Kassas \textit{et al.} \cite{kassas2019new} state that the SGP4 propagator has a positioning error of around 3 km while estimating satellite position. 
The Doppler shift observation model of UT for a straightforward approach is expressed as
\begin{align}
    \Dot{\rho}^\ell_{\sf UT}&=\left(\hat{\mathbf{v}}_{\sf sat}^{\ell}-\mathbf{v}_{\sf UT}\right)\cdot\frac{\hat{\mathbf{x}}_{\sf sat}^{\ell}-\mathbf{x}_{\sf UT}}{\|\hat{\mathbf{x}}_{\sf sat}^{\ell}-\mathbf{x}_{\sf UT}\|}+
     {\sf c}\big( a_{1,{\sf UT}}+a_{2,{\sf UT}}\left(t_{\sf R,UT} \right. \notag \\
     &\left. -t_{0,{\sf UT}}\right) 
    +\Dot{\psi}_{\sf UT}\left(t_{\sf R,UT}\right)\big)-{\sf c}\big(a^\ell_1+a^\ell_2\left(t^\ell_{\sf T}-t^\ell_0\right)  \notag \\
    &+\dot\psi^\ell\left(t^\ell_{\sf T}\right)\big) +\Delta{\Dot{T}}^{\ell}_{\sf UT}+\Delta{\Dot{I}}^{\ell}_{\sf UT}
    +\Delta{\Dot{\rho}}^\ell_{\sf UT}+\epsilon _{\Dot{\rho}^\ell_{\sf UT}},
    \label{eq: 8_delrhoUT}
\end{align}
where $\hat{\mathbf{x}}^{\ell}_{\sf sat}$ and $\hat{\mathbf{v}}^{\ell}_{\sf sat}$ are the estimated position and velocity of $\ell$-th satellite obtained using TLE and orbit propagator model.
$\Delta{\Dot{\rho}}^\ell_{\sf UT}= \left(\mathbf{v}_{\sf sat}^{\ell}-\mathbf{v}_{\sf UT}\right)\cdot\frac{\mathbf{x}_{\sf sat}^{\ell}-\mathbf{x}_{\sf UT}}{\|\mathbf{x}_{\sf sat}^{\ell}-\mathbf{x}_{\sf UT}\|}
-\left(\hat{\mathbf{v}}_{\sf sat}^{\ell}-\mathbf{v}_{\sf UT}\right)\cdot\frac{\hat{\mathbf{x}}_{\sf sat}^{\ell}-\mathbf{x}_{\sf UT}}{\|\hat{\mathbf{x}}_{\sf sat}^{\ell}-\mathbf{x}_{\sf UT}\|}$ is due to the ephemeris error of $\ell$-th LEO satellite in UT. Similarly, the observation model for the base receiver is expressed as
\begin{align}
    \Dot{\rho}^\ell_{\sf B} &= \hat{\mathbf{v}}_{\sf sat}^{\ell}\cdot\frac{\hat{\mathbf{x}}_{\sf sat}^{\ell}-\mathbf{x}_{\sf B}}{\|\hat{\mathbf{x}}_{\sf sat}^{\ell}-\mathbf{x}_{\sf B}\|}+
     {\sf c}\big(a_{1,{\sf B}}+a_{2,{\sf B}}\left(t_{\sf R,B}-t_{0,{\sf B}}\right)  \notag \\
    &+\dot\psi_{\sf B}(t_{\sf R,B})\big)
    -{\sf c}\big(a^\ell_1+a^\ell_2 
    \left(t^\ell_{\sf T}-t^\ell_0\right)
    +\dot\psi^\ell(t^\ell_{\sf T})\big) \notag \\
    &+\Delta{\Dot{T}}^{\ell}_{\sf B}+\Delta{\Dot{I}}^{\ell}_{\sf B} 
    +\Delta{\Dot{\rho}}^\ell_{\sf B}+\epsilon _{\Dot{\rho}^\ell_{\sf B}},
    \label{eq: 9_delrhoB}
\end{align} 
where $\mathbf{x}_{\sf B}$ and $\Delta{\Dot{\rho}}^\ell_{\sf B}= \mathbf{v}_{\sf sat}^{\ell}\cdot\frac{\mathbf{x}_{\sf sat}^{\ell}-\mathbf{x}_{\sf B}}{\|\mathbf{x}_{\sf sat}^{\ell}-\mathbf{x}_{\sf B}\|}
-\hat{\mathbf{v}}_{\sf sat}^{\ell}\cdot\frac{\hat{\mathbf{x}}_{\sf sat}^{\ell}-\mathbf{x}_{\sf B}}{\|\hat{\mathbf{x}}_{\sf sat}^{\ell}-\mathbf{x}_{\sf B}\|}$ are the known position of the base receiver and Doppler shift measurement error due to the ephemeris error of $\ell$-th LEO satellite in the base receiver, respectively.

The Doppler shift measurement consists of error terms like ionospheric error, troposphere error, ephemeris error, error related to clock, and measurement error. The ionospheric error and tropospheric error are curtailed by using standard models like Klobuchar model \cite{klabuchar} for the ionosphere and Saastamoinen model \cite{saastamoinen1972atmospheric} for the troposphere. 
However, errors related to satellite clock and ephemeris errors do not disappear.
The use of an additional base receiver with the UT can furnish errors related to satellite clock and ephemeris inaccuracies by subtracting the Doppler shift measurement of UT from the measurement of the base receiver. This difference in Doppler shift measurement between the base receiver and UT is considered a single-difference Doppler shift measurement.

Despite that, there is a clock synchronization error between the base receiver and UT. 
The Doppler shift measurement errors due to the satellite ephemeris error for the base receiver and UT are identical when the distance between them is relatively short.
However, the base receiver can not estimate the Doppler shift measurement error properly due to the ephemeris error, as the distance between the base receiver and UT increases significantly.
Moreover, the error terms related to clock drift, ephemeris, and atmospheric error still exist. 
To compensate for the error terms due to UT clock drift and other errors (ephemeris and atmospheric), the single-difference Doppler shift measurement is subtracted from a reference satellite measurement, which is considered a double-difference Doppler shift measurement. 
The measurement of the $l_0$-th satellite among the visible satellite measurements from the UT is considered as a reference satellite. 
To overcome the deficiency of positioning while increasing the distance between the base receiver and UT, the proposed ephemeris error correction algorithm recalculates the Doppler shift measurement error due to the ephemeris error of UT at the initial position estimated by the double-difference Doppler shift-based positioning algorithm.

\section{Robust double-difference Doppler shift-based positioning framework}
Herein, the proposed robust double-difference Doppler shift-based positioning (3DPose) framework is explained in detail. The proposed 3DPose framework comprises double-difference Doppler shift measurements and an ephemeris error correction algorithm, which is described below.

\subsection{Double-difference Doppler Shift measurement}

The Doppler shift observation model, for a straightforward approach, suffers from errors related to satellite clock, ephemeris, and atmospheric conditions that cause remarkable errors in positioning.
Hence, the double-difference Doppler shift approach is introduced, which considers Doppler shift measurements from the base receiver and UT. 
Firstly, to minimize the error terms, the single-difference Doppler shift is taken into account by subtracting the Doppler shift measurement of UT from the measurement of the base receiver. By utilizing the \eqref{eq: 8_delrhoUT} and \eqref{eq: 9_delrhoB}, the first step of double-difference Doppler shift for UT with $\ell$-th satellite is expressed as \cite{dif9843493}
\begin{align}
    \Dot{\rho}^{\ell}_{\sf UT,B} &= \Dot{\rho}^\ell_{\sf UT}-\Dot{\rho}^\ell_{\sf B} + 
    \hat{\mathbf{v}}_{\sf sat}^{\ell}\cdot\frac{\hat{\mathbf{x}}_{\sf sat}^{\ell}-\mathbf{x}_{\sf B}}{\|\hat{\mathbf{x}}_{\sf sat}^{\ell}-\mathbf{x}_{\sf B}\|} \notag \\
    &=\left(\hat{\mathbf{v}}_{\sf sat}^{\ell}-\mathbf{v}_{\sf UT}\right)\cdot\frac{\hat{\mathbf{x}}_{\sf sat}^{\ell}-\mathbf{x}_{\sf UT}}{\|\hat{\mathbf{x}}_{\sf sat}^{\ell}-\mathbf{x}_{\sf UT}\|}
     +{\sf c}\big(a_{1,{\sf UT}}+a_{2,{\sf UT}}\left(t_{\sf R,UT}-t_{0,{\sf UT}}\right) \notag \\
    & +\dot\psi_{\sf UT}\left(t_{\sf R,UT}\right)\big)
    -{\sf c}\left(a_{1,{\sf B}}+a_{2,{\sf B}}\left(t_{\sf R,B}-t_{0,{\sf B}}\right)+\dot\psi_{\sf B}(t_{\sf R,B})\right) \notag \\
    & +\Delta{\Dot{T}}^{\ell}_{\sf UT,B}+\Delta{\Dot{I}}^{\ell}_{\sf UT,B} 
    +\Delta{\Dot{\rho}}^{\ell}_{\sf UT,B}+\epsilon _{\Dot{\rho}^\ell_{\sf UT,B}},
    \label{eq: 10_singleD}
\end{align}
where $\Delta{\Dot{T}}^{\ell}_{\sf UT,B} = \Delta{\Dot{T}}^{\ell}_{\sf UT}-\Delta{\Dot{T}}^{\ell}_{\sf B}$, $\Delta{\Dot{I}}^{\ell}_{\sf UT,B} = \Delta{\Dot{I}}^{\ell}_{\sf UT}-\Delta{\Dot{I}}^{\ell}_{\sf B}$,
$\Delta{\Dot{\rho}}^{\ell}_{\sf UT,B} = \Delta{\Dot{\rho}}^{\ell}_{\sf UT}-\Delta{\Dot{\rho}}^{\ell}_{\sf B}$, and $\epsilon _{\Dot{\rho}^\ell_{\sf UT,B}}=\epsilon _{\Dot{\rho}^\ell_{\sf UT}}-\epsilon _{\Dot{\rho}^\ell_{\sf B}}$. However, error terms related to base receiver clock, UT clock, ephemeris, and atmospheric still exist. To compensate for the error terms due to base receiver and UT clock and other error terms (ephemeris and atmospheric), the single-difference Doppler shift measurement is subtracted from a reference satellite measurement and considered as a double-difference Doppler shift measurement. The measurement of the $l_0$-th satellite among the available measurements from the UT is considered as a reference satellite.
Similarly, as \eqref{eq: 10_singleD}, the first step of double-difference Doppler shift for UT with the reference $l_0$-th satellite is expressed as
\begin{align}
    \Dot{\rho}^{[\ell_0]}_{\sf UT,B} &= \Dot{\rho}^{[\ell_0]}_{\sf UT}-\Dot{\rho}^{[\ell_0]}_{\sf B} + 
    \hat{\mathbf{v}}_{\sf sat}^{[\ell_0]}\cdot\frac{\hat{\mathbf{x}}_{\sf sat}^{[\ell_0]}-\mathbf{x}_{\sf B}}{\|\hat{\mathbf{x}}_{\sf sat}^{[\ell_0]}-\mathbf{x}_{\sf B}\|} \notag \\
    &=\left(\hat{\mathbf{v}}_{\sf sat}^{[\ell_0]}-\mathbf{v}_{\sf UT}\right)\cdot\frac{\hat{\mathbf{x}}_{\sf sat}^{[\ell_0]}-\mathbf{x}_{\sf UT}}{\|\hat{\mathbf{x}}_{\sf sat}^{[\ell_0]}-\mathbf{x}_{\sf UT}\|}
     +{\sf c}\big(a_{1,{\sf UT}}+a_{2,{\sf UT}}\left(t_{\sf R,UT}-t_{0,{\sf UT}}\right) \notag \\
    & +\dot\psi_{\sf UT}\left(t_{\sf R,UT}\right)\big)
    -{\sf c}\left(a_{1,{\sf B}}+a_{2,{\sf B}}\left(t_{\sf R,B}-t_{0,{\sf B}}\right)+\dot\psi_{\sf B}(t_{\sf R,B})\right) \notag \\
    &+\Delta{\Dot{T}}^{[\ell_0]}_{\sf UT,B}+\Delta{\Dot{I}}^{[\ell_0]}_{\sf UT,B} 
    +\Delta{\Dot{\rho}}^{[\ell_0]}_{\sf UT,B}+\epsilon _{\Dot{\rho}^{[\ell_0]}_{\sf UT,B}}.
    \label{eq: 10_singleD_B}
\end{align}

The double-difference Doppler shift measurement for UT is {obtained by subtracting the $\ell$-th satellite measurement \eqref{eq: 10_singleD} from $\ell_0$-th satellite measurement \eqref{eq: 10_singleD_B}}, which is expressed as
\begin{align}
    \Dot{\rho}^{[\ell_0,\ell]}_{\sf UT,B}&= \Dot{\rho}^{[\ell_0]}_{\sf UT,B}-\Dot{\rho}^{\ell}_{\sf UT,B} \notag \\   
    &= \left(\hat{\mathbf{v}}_{\sf sat}^{[\ell_0]}-\mathbf{v}_{\sf UT}\right)\cdot\frac{\hat{\mathbf{x}}_{\sf sat}^{[\ell_0]}-\mathbf{x}_{\sf UT}}{\|\hat{\mathbf{x}}_{\sf sat}^{[\ell_0]}-\mathbf{x}_{\sf UT}\|}
    -\left(\hat{\mathbf{v}}_{\sf sat}^{\ell}-\mathbf{v}_{\sf UT}\right)\cdot\frac{\hat{\mathbf{x}}_{\sf sat}^{\ell}-\mathbf{x}_{\sf UT}}{\|\hat{\mathbf{x}}_{\sf sat}^{\ell}-\mathbf{x}_{\sf UT}\|}
      \notag \\
    &+\Delta{\Dot{T}}^{[\ell_0]}_{\sf UT,B}-\Delta{\Dot{T}}^{\ell}_{\sf UT,B}+\Delta{\Dot{I}}^{[\ell_0]}_{\sf UT,B} -\Delta{\Dot{I}}^{\ell}_{\sf UT,B} 
    +\Delta{\Dot{\rho}}^{[\ell_0]}_{\sf UT,B}-\Delta{\Dot{\rho}}^\ell_{\sf UT,B}+\epsilon _{\Dot{\rho}^{[\ell_0]}_{\sf UT,B}}-\epsilon _{\Dot{\rho}^\ell_{\sf UT,B}} \notag \\
    &= \left(\hat{\mathbf{v}}_{\sf sat}^{[\ell_0]}-\mathbf{v}_{\sf UT}\right)\cdot\frac{\hat{\mathbf{x}}_{\sf sat}^{[\ell_0]}-\mathbf{x}_{\sf UT}}{\|\hat{\mathbf{x}}_{\sf sat}^{[\ell_0]}-\mathbf{x}_{\sf UT}\|}
    - \left(\hat{\mathbf{v}}_{\sf sat}^{\ell}-\mathbf{v}_{\sf UT}\right)\cdot\frac{\hat{\mathbf{x}}_{\sf sat}^{\ell}-\mathbf{x}_{\sf UT}}{\|\hat{\mathbf{x}}_{\sf sat}^{\ell}-\mathbf{x}_{\sf UT}\|}  \notag \\
    &+\Delta{\Dot{T}}^{{[l_0,l]}}_{\sf UT,B}+\Delta{\Dot{I}}^{[\ell_0,\ell]}_{\sf UT,B}+
    \Delta{\Dot{\rho}}^{[\ell_0,\ell]}_{\sf UT,B}+\epsilon _{\Dot{\rho}^{[\ell_0,\ell]}_{\sf UT,B}}, 
    \label{eq: 11_dd}   
\end{align}
where $\Delta{\Dot{T}}^{[\ell_0,\ell]}_{\sf UT,B}=\Delta{\Dot{T}}^{[\ell_0]}_{\sf UT,B}-\Delta{\Dot{T}}^{\ell}_{\sf UT,B}$, $\Delta{\Dot{I}}^{[\ell_0,\ell]}_{\sf UT,B}=\Delta{\Dot{I}}^{[\ell_0]}_{\sf UT,B}-\Delta{\Dot{I}}^{\ell}_{\sf UT,B}$, and
$\Delta{\Dot{\rho}}^{[\ell_0,\ell]}_{\sf B}= \left(\mathbf{v}_{\sf sat}^{[\ell_0]}-\mathbf{v}_{\sf B}\right)\cdot\frac{\mathbf{x}_{\sf sat}^{[\ell_0]}-\mathbf{x}_{\sf B}}{\|\mathbf{x}_{\sf sat}^{[\ell_0]}-\mathbf{x}_{\sf B}\|}
-\left(\hat{\mathbf{v}}_{\sf sat}^{\ell}-\mathbf{v}_{\sf B}\right)\cdot\frac{\hat{\mathbf{x}}_{\sf sat}^{\ell}-\mathbf{x}_{\sf B}}{\|\hat{\mathbf{x}}_{\sf sat}^{\ell}-\mathbf{x}_{\sf B}\|}$\vspace{1.5mm} are the difference of tropospheric delay rate error, difference of ionospheric delay rate error, and difference of pseudornge rate, respectively.
With respect to unknowns ($\mathbf{x}_{\sf UT}$ and $\mathbf{v}_{\sf UT}$), \eqref{eq: 11_dd} is a non-linear equation. 
Considering the first-order Taylor approximation, the linearized pseudorange rate of $\ell$-th measurement from \eqref{eq: 11_dd} is expressed as
\begin{equation}
    \Dot{\rho}^{[\ell_0,\ell]}_{\sf UT,B}\approx \Dot{\rho}^{[\ell_0,\ell]}_{{\sf UT,B},0}+\nabla\Dot{\rho}^{[\ell_0,\ell]}_{{\sf UT,B},0}\cdot\Delta \mathbf{x}_{\sf pv}+\varepsilon^{[\ell_0,\ell]},
    \label{eq: 12_T}
\end{equation}
where $\Dot{\rho}^{[\ell_0,\ell]}_{{\sf UT,B},0}$, $\nabla\Dot{\rho}^{[\ell_0,\ell]}_{{\sf UT,B},0}$, and $\varepsilon^{[\ell_0,\ell]}$ are the difference in initial pseudorange rate estimation, partial derivative of difference pseudorange rate of $\ell$-th measurement considering an initial estimation $\mathbf{x}_{{\sf pv},0}$, and the total error term, respectively. 
$\Delta \mathbf{x}_{\sf pv}=\left[\Delta\mathbf{x}_{\sf UT} \hspace{.5em} \Delta\mathbf{v}_{\sf UT}\right]$ is the correction of UT position and velocity. The difference in initial pseudorange rate estimation is obtained as follows.
\begin{align}
    \Dot{\rho}^{[\ell_0,\ell]}_{{\sf UT,B},0} &= \left(\hat{\mathbf{v}}_{\sf sat}^{[\ell_0]}-\mathbf{v}_{{\sf UT},0}\right)\cdot\frac{\hat{\mathbf{x}}_{\sf sat}^{[\ell_0]}-\mathbf{x}_{{\sf UT},0}}{\|\hat{\mathbf{x}}_{\sf sat}^{[\ell_0]}-\mathbf{x}_{{\sf UT},0}\|}
    -\left(\hat{\mathbf{v}}_{\sf sat}^{\ell}-\mathbf{v}_{{\sf UT},0}\right)\cdot\frac{\hat{\mathbf{x}}_{\sf sat}^{\ell}-\mathbf{x}_{{\sf UT},0}}{\|\hat{\mathbf{x}}_{\sf sat}^{\ell}-\mathbf{x}_{{\sf UT},0}\|}.
    \label{eq: initialRohDot}
\end{align}

The measurements for $L$ satellites yield 
\begin{align}
    \Delta\Dot{\mathbf{\uprho}}= 
    \begin{bmatrix}
        \Dot{\rho}^{[\ell_0,1]}_{\sf UT,B}-\Dot{\rho}^{[\ell_0,1]}_{{\sf UT,B},0}\\
        \Dot{\rho}^{[\ell_0,2]}_{\sf UT,B}-\Dot{\rho}^{[\ell_0,2]}_{{\sf UT,B},0}\\
        \vdots\\
        \Dot{\rho}^{[\ell_0,L-1]}_{\sf UT,B}-\Dot{\rho}^{[\ell_0,L-1]}_{{\sf UT,B},0}
    \end{bmatrix} &= 
    \begin{bmatrix}
        \left(\nabla\Dot{\rho}^{[\ell_0,1]}_{{\sf UT,B},0}\right)^T\\
        \left(\nabla\Dot{\rho}^{[\ell_0,2]}_{{\sf UT,B},0}\right)^T\\
        \vdots\\
        \left(\nabla\Dot{\rho}^{[\ell_0,L-1]}_{{\sf UT,B},0}\right)^T\\
    \end{bmatrix}
    \Delta \mathbf{x}_{\sf pv}+\mathbf{\upvarepsilon}  \notag \\
    \Leftrightarrow \Delta\Dot{\mathbf{\uprho}}&= \mathbf{G}\Delta \mathbf{x}_{\sf pv}+\mathbf{\upvarepsilon},
    \label{eq: 13_dd_matrix}
\end{align}
where $\mathbf{G}=[\nabla\Dot{\rho}^{[\ell_0,1]}_{{\sf UT,B},0} \nabla\Dot{\rho}^{[\ell_0,2]}_{{\sf UT,B},0} \hdots \nabla\Dot{\rho}^{[\ell_0,L-1]}_{{\sf UT,B},0}]^{T}$\vspace{0.08cm} is the Jacobian matrix which is expressed in \eqref{eq: 14_Gmatrix_B}.
\begin{figure}[b]
\begin{align}
\hline
    \mathbf{G}=
    \begin{bmatrix}
        g^{[\ell_0]}_{x}-g^{1}_{x}&g^{[\ell_0]}_{y}-g^{1}_{y}&g^{[\ell_0]}_{z}-g^{1}_{z}& -e^{[\ell_0]}_{x}+e^{1}_{x}& -e^{[\ell_0]}_{y}+e^{1}_{y}& -e^{[\ell_0]}_{z}+e^{1}_{z}\\[.2cm]
       g^{[\ell_0]}_{x}-g^{2}_{x}&g^{[\ell_0]}_{y}-g^{2}_{y}&g^{[\ell_0]}_{z}-g^{2}_{z}& -e^{[\ell_0]}_{x}+e^{2}_{x}& -e^{[\ell_0]}_{y}+e^{2}_{y}& -e^{[\ell_0]}_{z}+e^{2}_{z}\\
        \vdots & \vdots & \vdots & \vdots & \vdots & \vdots \\
        g^{[\ell_0]}_{x}-g^{L-1}_{x}&g^{[\ell_0]}_{y}-g^{L-1}_{y}&g^{[\ell_0]}_{z}-g^{L-1}_{z}& -e^{[\ell_0]}_{x}+e^{L-1}_{x}& -e^{[\ell_0]}_{y}+e^{L-1}_{y}& -e^{[\ell_0]}_{z}+e^{L-1}_{z}
    \end{bmatrix},
    \label{eq: 14_Gmatrix_B}
\end{align}
\end{figure}
where $g^\ell_{x}=\frac{\Delta v^\ell_{x}}{\rho^\ell}(({e^\ell_x})^{2}-1)+\frac{e^\ell_{x}}{\rho^\ell}(\Delta v^\ell_{y}e^\ell_{y}+\Delta v^\ell_{z}e^\ell_{z})$, $g^\ell_{y}=\frac{\Delta v^\ell_{y}}{\rho^\ell}((e^\ell_y)^2-1)+\frac{e^\ell_{y}}{\rho^\ell}(\Delta v^\ell_{x}e^\ell_{x}+\Delta v^\ell_{z}e^\ell_{z})$,\vspace{0.08cm} 
$g^\ell_{z}=\frac{\Delta v^\ell_{z}}{\rho^\ell}((e^\ell_z)^2-1)+\frac{e^\ell_{z}}{\rho^\ell}(\Delta v^\ell_{x}e^\ell_{x}+\Delta v^\ell_{y}e^\ell_{y})$, 
$\rho^\ell= \|\hat{\mathbf{x}}^\ell_{\sf sat}-\mathbf{x}_{{\sf UT},0}\|$,\vspace{0.08cm} and
$[\Delta v^\ell_{x}\hspace{.2cm} \Delta v^\ell_{y} \hspace{.2cm} \Delta v^\ell_{z}]^T = \hat{\mathbf{v}}^\ell_{\sf sat}-\mathbf{v}_{{\sf UT},0}$; 
$[e^\ell_{x} \hspace{.2cm} e^\ell_{y} \hspace{.2cm} e^\ell_{z}]^{T} = (\hat{\mathbf{x}}^\ell_{\sf sat}-\mathbf{x}_{{\sf UT},0})/\rho^\ell$, is the LOS unit vector between $\ell$-th satellite and initial estimation.
The least squares (LS) solution of \eqref{eq: 13_dd_matrix} is
\begin{equation}
    \Delta \mathbf{x}_{\sf pv}=\left(\mathbf{G}^{T}\mathbf{G}\right)^{-1} \mathbf{G}^{T}\Delta\Dot{\mathbf{\uprho}}.
    \label{eq: 15_LS}
\end{equation}
The LS solution treats all the Doppler shift measurements as equal importance.
In the practical environment, the Doppler shift measurements are affected due to several factors, including the relative geometry between satellite and UT as well as the probability of induced errors. Therefore, all the Doppler shift measurements do not contribute equally while utilizing positioning solutions. 
The quality of {the} Doppler shift measurements can be indicated by the ratio of signal power and noise power, defined as the signal-to-noise ratio (SNR) of the received signal \cite{10297313}. 
{Higher SNR contributes less Doppler shift measurement error.}
In light of this, SNR is used as a weighting factor in the weighted least squares (WLS) solution {\cite{vincent2020doppler}.}
{Other weighting factors, such as low election angle, cause a larger signal propagation path, reducing the signal strength, i.e., less SNR.}
Using weighting factors the WLS solution of \eqref{eq: 13_dd_matrix} is
\begin{equation}
    \Delta \mathbf{x}_{\sf pv}=\left(\mathbf{G}^{T}\mathbf{W}\mathbf{G}\right)^{-1} \mathbf{G}^{T}\mathbf{W}\Delta\Dot{\mathbf{\uprho}},
    \label{eq: 15_WLS}
\end{equation}
where $\mathbf{W}$ is the weight matrix.
The corrected position is obtained by adding correction ($\Delta \mathbf{x}_{\sf pv}$) with the initial estimation ($\mathbf{x}_{{\sf pv},0}$) as follows:
\begin{equation}
    \mathbf{x}_{{\sf pv},1}=\mathbf{x}_{{\sf pv},0}+\Delta \mathbf{x}_{\sf pv}.
    \label{eq: 16_estimation_x_1}
\end{equation}

The position ($\mathbf{x}_{{\sf pv},1}$) obtained from \eqref{eq: 16_estimation_x_1} is considered new initial estimation ($\mathbf{x}_{{\sf pv},0}$) for \eqref{eq: 12_T} and subsequently, all the measurements for $L$ satellites are updated using \eqref{eq: 13_dd_matrix}. 
Then $\Delta \mathbf{x}_{\sf pv}$ is calculated using \eqref{eq: 15_LS} for getting LS solution or using \eqref{eq: 15_WLS} for WLS solution. The optimal position is obtained iteratively by adding correction ($\Delta \mathbf{x}_{\sf pv}$) with the previous estimation until $\Delta \mathbf{x}_{\sf pv}$ is less than a threshold value ($\zeta$) as follows:
\begin{equation}
    \hat{\mathbf{x}}_{{\sf pv},k}=\mathbf{x}_{{\sf pv},k-1}+\Delta \mathbf{x}_{\sf pv},
    \label{eq: 16_estimation_x}
\end{equation}
where $k$ is the number of iterations.

\subsection{Ephemeris Error Correction Algorithm}
\begin{figure}[t]
    \centering
    \includegraphics[width=0.9\linewidth]{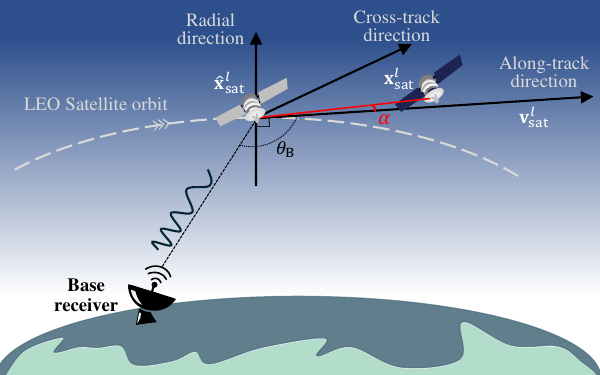}
    \caption{Overview of satellite position error in the radial, cross-track, and along-track direction.}
    \label{fig:alongtrack}
\end{figure}

The above-mentioned Doppler shift-based method can not effectively minimize the Doppler shift measurement error of UT due to the ephemeris error in the case of a long baseline.
The Doppler shift depends on the relative velocity between the receiver and satellite in the LOS direction. With the increase in the distance between the base receiver and UT, the LOS direction between the satellite and receiver also changes for the base receiver and UT. 
{Due to the change in LOS direction, the correction of the Doppler shift measurement error estimated by the base receiver is not equal to that of UT.}
It is necessary to model the Doppler shift measurement error of UT due to the satellite ephemeris error in the estimated UT position. 
The initial estimated position of UT is obtained using \eqref{eq: 16_estimation_x}. 
The Doppler shift measurement for the base receiver is 
\begin{align}
        \dot{\rho}_{\sf B}^\ell &= \mathbf{v}_{\sf sat}^\ell \cdot \frac{\mathbf{x}_{\sf sat}^\ell - \mathbf{x}_{\sf B}}{\| \mathbf{x}_{\sf sat}^\ell - \mathbf{x}_{\sf B} \|}+\hat{\epsilon}_{\sf B} \notag \\
        &=\mathbf{v}_{\sf sat}^\ell \cdot \frac{\hat{\mathbf{x}}_{\sf sat}^\ell + \Delta \mathbf{x}_{\sf sat}^\ell -\mathbf{x}_{\sf B}}{\| \mathbf{x}_{\sf sat}^\ell - \mathbf{x}_{\sf B} \|}+\hat{\epsilon}_{\sf B} \notag \\
    &=\mathbf{v}_{\sf sat}^\ell \cdot \frac{\hat{\mathbf{x}}_{\sf sat}^\ell - \mathbf{x}_{\sf B}}{\| \hat{\mathbf{x}}_{\sf sat}^\ell - \mathbf{x}_{\sf B} \|} \frac{\| \hat{\mathbf{x}}_{\sf sat}^\ell - \mathbf{x}_{\sf B} \|}{\| \mathbf{x}_{\sf sat}^\ell - \mathbf{x}_{\sf B} \|}+\mathbf{v}_{\sf sat}^\ell \cdot \frac{\Delta \mathbf{x}_{\sf sat}^\ell}{\| \mathbf{x}_{\sf sat}^\ell - \mathbf{x}_{\sf B} \|}+\hat{\epsilon}_{\sf B} \notag \\
        &= \hat{\dot{\rho}}_{\sf B}^\ell \frac{\hat{r}_{\sf B}^\ell}{r_{\sf B}^\ell} + \frac{\| \mathbf{v}_{\sf sat}^\ell \| \| \Delta \mathbf{x}_{\sf sat}^\ell \|\cos\alpha}{r_{\sf B}^\ell}+\hat{\epsilon}_{\sf B},
        \label{eq:baseDoppler1}
\end{align}
where $\| \Delta \mathbf{x}_{\sf sat}^\ell \| = \| \mathbf{x}_{\sf sat}^\ell - \hat{\mathbf{x}}_{\sf sat}^\ell \|$, $r_{\sf B}^\ell = \| \mathbf{x}_{\sf sat}^\ell - \mathbf{x}_{\sf B} \|$, $\hat{r}_{\sf B}^\ell = \| \hat{\mathbf{x}}_{\sf sat}^\ell - \mathbf{x}_{\sf B} \|$, $\hat{\dot{\rho}}_{\sf B}^\ell = \mathbf{v}_{\sf sat}^\ell\cdot \frac{\hat{\mathbf{x}}_{\sf sat}^\ell - \mathbf{x}_{\sf B}}{\| \hat{\mathbf{x}}_{\sf sat}^\ell - \mathbf{x}_{\sf B} \|}$, $\alpha$, and $\hat{\epsilon}_{\sf B}$ are the satellite position error, range between base receiver and satellite true position, estimated range between base receiver and satellite estimated position, estimated pseudorange rate in base receiver, angle between satellite position error vector and satellite velocity direction, and total error, respectively. 
The position error of the LEO satellite mainly condenses in the along-track direction \cite{zhao2023analysis,10288195} as shown in Fig~\ref{fig:alongtrack}.
Due to the lower altitude of the LEO satellite, the orbit of the LEO satellite is nearly circular, and the along-track direction is the tangential direction. Owing to this, the angle between the satellite position error vector and satellite velocity is almost zero \cite{10288195}. Substituting $\alpha=0$ in \eqref{eq:baseDoppler1}, we get 
\begin{equation}
    \dot{\rho}_{\sf B}^\ell = \hat{\dot{\rho}}_{\sf B}^\ell \frac{\hat{r}_{\sf B}^\ell}{r_{\sf B}^\ell} + \frac{\|\mathbf{v}_{\sf sat}^\ell \| \| \Delta \mathbf{x}_{\sf sat}^\ell \|}{r_{\sf B}^\ell}+\hat{\epsilon}_{\sf B}.
    \label{eq:baseDoppler}
\end{equation}

The satellite true position $\mathbf{x}_{\sf sat}^\ell$ is unknown to UT, and estimated position $\hat{\mathbf{x}}_{\sf sat}^\ell$ using TLE file and SGP4 orbit propagation model has errors. Considering the first-order Taylor approximation, the true range between the satellite and base receiver is as follows
\begin{align}
     \| \mathbf{x}_{\sf sat}^\ell - \mathbf{x}_{\sf B} \| &= \|\hat{\mathbf{x}}_{\sf sat}^\ell - \mathbf{x}_{\sf B} \| +\frac{\partial}{\partial \mathbf{x}} \left( \|\hat{\mathbf{x}}_{\sf sat}^\ell - \mathbf{x}_{\sf B} \| \right) \cdot \Delta \mathbf{x}_{\sf sat}^\ell \notag \\
     &= \|\hat{\mathbf{x}}_{\sf sat}^\ell - \mathbf{x}_{\sf B} \| + \frac{\hat{\mathbf{x}}_{\sf sat}^\ell - \mathbf{x}_{\sf B}}{\|\hat{\mathbf{x}}_{\sf sat}^\ell - \mathbf{x}_{\sf B} \|} \cdot \Delta \mathbf{x}_{\sf sat}^\ell \notag \\
    \Leftrightarrow r_{\sf B}^\ell &= \hat{r}_{\sf B}^\ell + \| \Delta \mathbf{x}_{\sf sat}^\ell \| \cos \theta_{\sf B},
     \label{eq:baseTaylor1}
\end{align}
where $\theta_{\sf B}$ is the angle between the LOS direction of the satellite to the base receiver and the satellite position error vector. As $\alpha=0$, $\theta_{\sf B}$ can also be defined as the angle between the LOS direction of the satellite to the base receiver and the satellite velocity direction. The measurement pseudorange rate is the component of satellite velocity in the LOS direction from satellite to base receiver. Thus, substituting $\cos\theta_{\sf B}=(\dot{\rho}^\ell_{\sf B}-\hat{\epsilon}_{\sf B})/\|\mathbf{v}^\ell_{\sf sat}\|$ in \eqref{eq:baseTaylor1}, we get 
\begin{equation}
     r_{\sf B}^\ell = \hat{r}_{\sf B}^\ell + \| \Delta \mathbf{x}_{\sf sat}^\ell \| \frac{\dot{\rho}^\ell_{\sf B} -\hat{\epsilon}_{\sf B}}{\|\mathbf{v}^\ell_{\sf sat}\|}.
\end{equation}

\begin{figure*}[ht]
    \centering
    \includegraphics[width=1\linewidth]{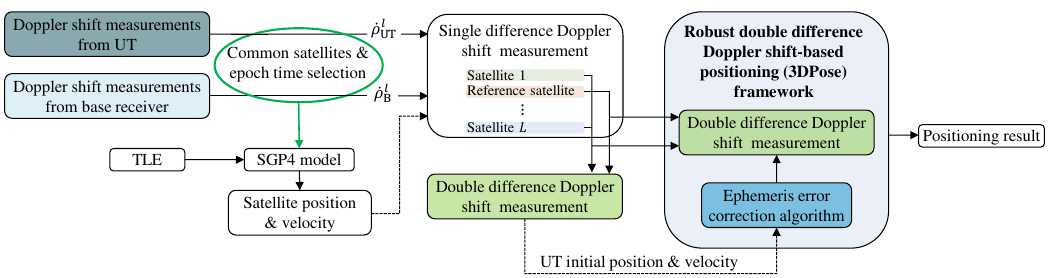}
    \caption{Block diagram of proposed 3DPose framework.}
    \label{fig: flowchart}
\end{figure*}


\begin{algorithm}[t]
    \caption{Ephemeris Error Correction} \label{alg: error}
    \KwInput{$\hat{\mathbf{x}}^\ell_{\sf sat}, \hat{\mathbf{v}}^\ell_{\sf sat}, \mathbf{x}_{\sf B}, f^\ell_{\sf d,B}, f^\ell_{\sf d,UT}, \lambda^\ell_T$, $\hat{\mathbf{x}}_{\sf UT}, \hat{\mathbf{v}}_{\sf UT}$, $L$} 
    $\Dot{\rho}^\ell_{\sf B}=f^\ell_{\sf d,B} \lambda^\ell_T, \Dot{\rho}^\ell_{\sf UT}=f^\ell_{\sf d,UT} \lambda^\ell_T$ 
    {\algorithmiccomment{Pseudorange rate for the base receiver and UT}} \\
    {Initialize: $\hat{\mathbf{x}}_{\sf UT}, \hat{\mathbf{v}}_{\sf UT},l$} \\
    \Repeat{$\ell=L$}
        {
       $\hat{r}_{\sf UT}^\ell = \| \hat{\mathbf{x}}_{\sf sat}^\ell - \hat{\mathbf{x}}_{\sf UT} \|$ 
       {\algorithmiccomment{Estimated range b/w base receiver and $\ell$-th satellite estimated position}}\\
        $\hat{\dot{\rho}}_{\sf UT}^\ell = \left( \hat{\mathbf{v}}_{\sf sat}^\ell - \hat{\mathbf{v}}_{\sf UT} \right) \cdot \frac{\hat{\mathbf{x}}_{\sf sat}^\ell - \hat{\mathbf{x}}_{\sf UT}}{\| \hat{\mathbf{x}}_{\sf sat}^\ell - \hat{\mathbf{x}}_{\sf UT} \|}$ {\algorithmiccomment{Estimated pseudorange rate of UT}}\\
        $\hat{r}_{\sf B}^\ell = \| \hat{\mathbf{x}}_{\sf sat}^\ell - \mathbf{x}_{\sf B} \|$ {\algorithmiccomment{Estimated range b/w UT and $\ell$-th satellite estimated position}}\\
        $\hat{\dot{\rho}}_{\sf B}^\ell = \hat{\mathbf{v}}_{\sf sat}^\ell \cdot \frac{\hat{\mathbf{x}}_{\sf sat}^\ell - \mathbf{x}_{\sf B}}{\| \hat{\mathbf{x}}_{\sf sat}^\ell - \mathbf{x}_{\sf B} \|}$  {\algorithmiccomment{Estimated pseudorange rate of base receiver}}\\ [5pt]
        \textnormal{Calculate} $\|\Delta \mathbf{x}^\ell_{\sf sat}\|$ \textnormal{using} \eqref{eq:deltaXsat} {\algorithmiccomment{Calculation of $\ell$-th satellite position error}}\\[2pt]
        \textnormal{Calculate} $ r^\ell_{\sf UT}$ \textnormal{using} \eqref{eq:rlUT} {\algorithmiccomment{Calculation of true range b/w the $\ell$-th satellite and UT }}\\[2pt]
        \textnormal{Calculate} $\Delta \Dot{\rho}^\ell_{\sf UT}$ \textnormal{using} \eqref{eq:delUT} {\algorithmiccomment{Calculation of pseudorange rate error due to ephemeris error}} \\
        $\ell\leftarrow \ell+1${\algorithmiccomment{Increament of satellite index}}
        }
        \KwOutput{$\Delta \Dot{\rho}^\ell_{\sf UT}$}
\end{algorithm}

Similarly, the true range between the satellite and UT is as follows 
\begin{equation}
     r_{\sf UT}^\ell = \hat{r}_{\sf Ut}^\ell + \| \Delta \mathbf{x}_{\sf sat}^\ell \| \frac{\dot{\rho}_{\sf UT}^\ell-\hat{\epsilon}_{\sf UT}}{\| \mathbf{v}_{\sf sat}^\ell \|},
     \label{eq:rlUT}
\end{equation}
where $r_{\sf UT}^\ell = \| \mathbf{x}_{\sf sat}^\ell - \hat{\mathbf{x}}_{\sf UT} \|$, 
$\hat{r}_{\sf UT}^\ell = \| \hat{\mathbf{x}}_{\sf sat}^\ell - \hat{\mathbf{x}}_{\sf UT} \|$, and $\hat{\epsilon}_{\sf UT}$
are the range between UT and satellite true position, estimated range between UT and satellite estimated position, and total error, respectively.
Substituting the value of $r_{\sf B}^\ell$ into \eqref{eq:baseDoppler} and rearranging we get
\begin{equation}
    \| \Delta \mathbf{x}_{\sf sat}^\ell \| = \frac{\hat{\dot{\rho}}_{\sf B}^\ell \hat{r}_{\sf B}^\ell \| \mathbf{v}_{\sf sat}^\ell \| - (\dot{\rho}_{\sf B}^\ell-\hat{\epsilon}_{\sf B}) \hat{r}_{\sf B}^\ell \| \mathbf{v}_{\sf sat}^\ell \|}{\left( \dot{\rho}_{\sf B}^\ell-\hat{\epsilon}_{\sf B} \right)^2 - \| \mathbf{v}_{\sf sat}^\ell \|}.
    \label{eq:deltaXsat}
\end{equation}

The pseudorange rate error due to the ephemeris error of UT in the estimated position ($\hat{\mathbf{x}}_{\sf UT}$) is as follows
\begin{align}
    \Delta \Dot{\rho}_{\sf UT}^\ell&=\left(\mathbf{v}_{\sf sat}^\ell - \hat{\mathbf{v}}_{\sf UT} \right) \cdot \frac{\mathbf{x}_{\sf sat}^\ell - \hat{\mathbf{x}}_{\sf UT}}{\| \mathbf{x}_{\sf sat}^\ell - \hat{\mathbf{x}}_{\sf UT} \|} - 
    \left( \mathbf{v}_{\sf sat}^\ell - \hat{\mathbf{v}}_{\sf UT} \right) \cdot \frac{\hat{\mathbf{x}}_{\sf sat}^\ell - \hat{\mathbf{x}}_{\sf UT}}{\| \hat{\mathbf{x}}_{\sf sat}^\ell -\hat{\mathbf{x}}_{\sf UT} \|} \notag \\ 
    &= \left( \mathbf{v}_{\sf sat}^\ell -\hat{\mathbf{v}}_{\sf UT} \right) \cdot \frac{\hat{\mathbf{x}}_{\sf sat}^\ell + \Delta \mathbf{x}_{\sf sat}^\ell -\hat{\mathbf{x}}_{\sf UT}}{\| \mathbf{x}_{\sf sat}^\ell - \hat{\mathbf{x}}_{\sf UT} \|} - 
    \hat{\dot{\rho}}_{\sf UT}^\ell \notag \\
    &= \left( \mathbf{v}_{\sf sat}^\ell - \hat{\mathbf{v}}_{\sf UT} \right) \cdot \frac{\hat{\mathbf{x}}_{\sf sat}^\ell - \hat{\mathbf{x}}_{\sf UT}}{\| \hat{\mathbf{x}}_{\sf sat}^\ell - \hat{\mathbf{x}}_{\sf UT} \|} \frac{\| \hat{\mathbf{x}}_{\sf sat}^\ell - \hat{\mathbf{x}}_{\sf UT} \|}{\| \mathbf{x}_{\sf sat}^\ell - \hat{\mathbf{x}}_{\sf UT} \|} 
    + \left( \mathbf{v}_{\sf sat}^\ell - \hat{\mathbf{v}}_{\sf UT} \right) \cdot \frac{\Delta \mathbf{x}_{\sf sat}^\ell}{\| \mathbf{x}_{\sf sat}^\ell - \hat{\mathbf{x}}_{\sf UT} \|} 
    - \hat{\dot{\rho}}_{\sf UT}^\ell \notag \\
    &= \hat{\dot{\rho}}_{\sf UT}^\ell \frac{\hat{r}_{\sf UT}^\ell}{r_{\sf UT}^\ell} + \frac{\| \mathbf{v}_{\sf sat}^\ell - \hat{\mathbf{v}}_{\sf UT} \| \| \Delta \mathbf{x}_{\sf sat}^\ell\|\cos\alpha}{r_{\sf UT}^\ell}- \hat{\dot{\rho}}_{\sf UT}^\ell,
    \label{eq:delUT1} 
\end{align}
where $\hat{\dot{\rho}}_{\sf UT}^\ell = \left( \mathbf{v}_{\sf sat}^\ell - \hat{\mathbf{v}}_{\sf UT} \right) \cdot \frac{\hat{\mathbf{x}}_{\sf sat}^\ell - \hat{\mathbf{x}}_{\sf UT}}{\| \hat{\mathbf{x}}_{\sf sat}^\ell - \hat{\mathbf{x}}_{\sf UT} \|}$
is estimated pseudorange rate in UT.
As the angle between relative velocity and satellite position error, $\alpha\approx0$, substituting $\cos\alpha=1$ in \eqref{eq:delUT1} and after rearranging, we get
\begin{equation}
     \Delta \Dot{\rho}_{\sf UT}^\ell= \hat{\dot{\rho}}_{\sf UT}^\ell \left(\frac{\hat{r}_{\sf UT}^\ell}{r_{\sf UT}^\ell} - 1\right) + \frac{\| \mathbf{v}_{\sf sat}^\ell - \hat{\mathbf{v}}_{\sf UT} \| \| \Delta \mathbf{x}_{\sf sat}^\ell \|}{r_{\sf UT}^\ell}.
     \label{eq:delUT} 
\end{equation}

Substituting the value of $\| \Delta \mathbf{x}_{\sf sat}^\ell \|$ and $r_{\sf UT}^\ell$ in \eqref{eq:delUT}, we can recalculate the Doppler shift measurement error of UT in the initial estimated position. The proposed ephemeris error correction algorithm is summarized in Algorithm~\ref{alg: error}.
This measurement error is used to cancel the term $\Delta\dot\rho^\ell_{\sf UT, B}$ from \eqref{eq: 10_singleD}. As a result the term $ \Delta{\Dot{\rho}}^{[\ell_0,\ell]}_{\sf UT,B}$ is eliminated from \eqref{eq: 11_dd}. After that, \eqref{eq: initialRohDot}, \eqref{eq: 13_dd_matrix}, \eqref{eq: 15_WLS}, \eqref{eq: 16_estimation_x_1}, and \eqref{eq: 16_estimation_x} are used to estimate the UT position more precisely. The overall block diagram of our proposed 3DPose framework is shown in Fig~\ref{fig: flowchart}.

\begin{algorithm}[t]
    \caption{Proposed 3DPose Framework} \label{alg:3DPose_WLS}
        \KwInt{$\mathbf{x}_{{\sf pv},0}, \ell\leftarrow1$}
    \Repeat{$\ell=L$}
        {\textnormal{Calculate} $\Delta \Dot{\rho}^\ell_{\sf UT}$ \textnormal{using Algorithm~\ref{alg: error}} \\
        \If{$\ell=1$}{
            $\Dot{\rho}^{[\ell_0]}_{\sf UT,B}= \Dot{\rho}^{[\ell_0]}_{\sf UT}-\Delta \Dot{\rho}^{[\ell_0]}_{\sf UT}$ \\ 
            \Continue }
        \Else{
            $\Dot{\rho}^{\ell}_{\sf UT,B}= \Dot{\rho}^\ell_{\sf UT}-\Delta \Dot{\rho}^\ell_{\sf UT}$
        }
        $\Dot{\rho}^{[\ell_0,\ell]}_{\sf UT,B}= \Dot{\rho}^{[\ell_0]}_{\sf UT,B}-\Dot{\rho}^{\ell}_{\sf UT,B}$ \\
         $\ell\leftarrow \ell+1$} 
     \Repeat{$\|\Delta \mathbf{x}_{\sf pv}\|<\zeta$}
         {
         \textnormal{Calculate} $\Dot{\rho}^{[\ell_0,\ell]}_{{\sf UT,B},0}$ \textnormal{using} \eqref{eq: initialRohDot}\\
         \begin{case}
             $\mathbf{W}=\mathbf{I}$ \\
          \end{case}
          \begin{case}
             $\mathbf{W}={\sf diag}\left(\frac{{\sf max}(\mathbf{s})-\mathbf{s}}{{\sf max}(\mathbf{s})-{\sf min}(\mathbf{s})}\right)$ \\
          \end{case}
         $\Delta \mathbf{x}_{\sf pv}=\left(\mathbf{G}^{T}\mathbf{W}\mathbf{G}\right)^{-1} \mathbf{G}^{T}\mathbf{W}\Delta\Dot{\mathbf{\uprho}}$ \\
         $\hat{\mathbf{x}}_{{\sf pv},k}=\mathbf{x}_{{\sf pv},k-1}+\Delta \mathbf{x}_{\sf pv}$ \\
         $k\leftarrow k+1$
         }
         \KwOutput{$\hat{\mathbf{x}}_{{\sf pv},k}$} 
\end{algorithm}

\begin{remark}  {\normalfont
    \textbf{(Clock synchronization deficiency of the base receiver and UT)}:}
    \label{remark:clock}
Unlike GNSS, the Clock used in LEO satellites are not synchronized with each other. For simplicity, existing literature \cite{shi2023revisiting,BENZERROUK2019496,8682554} considers the same satellite clock drift for all satellites, which is not practical. 
Using a base receiver with UT can eliminate the clock synchronization issue of LEO satellites. However, the lack of clock synchronization between the base receiver and UT needs to be addressed to reduce Doppler shift measurement error. 
Prior research has not considered the lack of clock synchronization between the base receiver and UT. 
Our proposed 3DPose framework uses double-difference Doppler shift measurement to address the clock synchronization issue between the base receiver and UT.
\end{remark}

\begin{remark} {\normalfont 
    \textbf{(Key difference from the prior differential Doppler positioning methods \cite{dif9843493})}:} \label{remark:longbaeline}
    The conventional differential Doppler positioning methods \cite{dif9843493} calculate Doppler shift measurement error due to the ephemeris error for the base receiver. This measurement error is used to compensate for the effect of the ephemeris error in UT positioning. This measurement error is similar for the base receiver and UT, while the baseline is short. With the increase in the baseline, Doppler shift measurement error due to erroneous ephemeris data changes for the base receiver and UT. In stark contrast to the previous study, we characterize and correct the Doppler shift measurement error due to the ephemeris error for the UT using the ephemeris error correction algorithm irrespective of the baseline length.
\end{remark}

\section{Result and Discussion}
\label{sec: Result_and_Discussion}
This section presents the comparison of positioning results in the case of the three variants of the proposed framework (Vanilla double-difference based on WLS, 3DPose based on LS, and 3DPose based on WLS) with the existing differential Doppler positioning method \cite{dif9843493} in terms of root mean square error (RMSE) in the North, East, Up (NEU) coordinate, time series positioning error analysis in NEU coordinate and \mbox{3-D} {positioning} error, and evaluation map for all the three scenarios.

\begin{itemize}
   {\item \textbf{Vanilla double-difference based on WLS}: The Vanilla double-difference is the proposed 3DPose framework considering the double-difference Doppler shift measurement without the ephemeris error correction algorithm to show the effect of removing the clock synchronization issue between the base receiver and UT.
   }
   
   {\item \textbf{3DPose based on LS}: This method utilizes double-difference Doppler Shift measurement and ephemeris error correction algorithm as explained in Algorithm~\ref{alg:3DPose_WLS} with case 1.
   }

   {\item \textbf{3DPose based on WLS}: Algorithm~\ref{alg:3DPose_WLS} with case 2 shows the detailed procedure of 3DPose based on WLS for getting robust positioning solution.
   }
\end{itemize}

The environments are created by considering the base receiver at Ajou University, Suwon, South Korea {for simulation evaluation}. 
The latitude and longitude of the base receiver are $37.282268^\circ$ north and $127.043524^\circ$ east, respectively. 
We consider a moving UT for position estimation. Three different scenarios are considered with three separate baselines.
{In each scenario, different urban driving trajectories are created utilizing Google Earth Pro software \cite{britt2019introduction}.}
{The simulation device is equipped with an Intel(R) Core(TM) i7-12700K @ 3610 MHz CPU and 32 GB RAM with 64-bit Windows 11.}

\begin{itemize}
    {\item \textbf{Scenario-I}: Scenario-I consists of a driving trajectory of around $\num{3.72}$ km where the distance between the base receiver and UT varies from $\num{1.02}$ km to $\num{2.48}$ km. It is considered a short baseline.
    }

    {\item \textbf{Scenario-II}: A driving trajectory of around $\num{8.10}$ km is considered in scenario-II, where the distance between the base receiver and UT changes from $\num{5.27}$ km to $\num{12.28}$ km. It is denoted as a medium baseline. 
    }

    {\item \textbf{Scenario-III}: Scenario-III utilizes a driving trajectory of around $\num{7.12}$ km where the distance betTeween the base receiver and UT changes from $\num{33.54}$ km to $\num{35.76}$ km. It is considered a long baseline.
    }
\end{itemize}

The Doppler shift is calculated by using MATLAB Satellite Communications Toolbox \cite{satCtoolbox} {with MATLAB version R2023a} and Starlink satellite's TLE files \cite{TLElink}.
In this {study}, we consider $118$ Starlink satellite's TLE files 
{with an altitude of 550 km and inclination of $53.21^\circ$ and $53.05^\circ$.}
{The TLE file contains the inclination angle, eccentricity, right ascension of the ascending node (RAAN), and argument of perigee of the satellite orbit. It also contains information on the mean anomaly and mean motion of the satellite at a certain time.}
{The constellation design parameters such as inclination, eccentricity, and altitude of $15$ different Starlink satellites among $118$ satellites are shown in Table~\ref{tab:TLE_table}.}


\begin{table}[t]\renewcommand{\arraystretch}{1}
    \caption{{Orbital parameter of different TLE files [43]}}
    \centering
    \begin{tabular}{|c|c|c|c|}
        \hline
         Satellite name & Inclination (degree) & Eccentricity & Altitude (km)  \\
         \hline
        STARLINK-1403 & $53.0524$ & $0.0001076$ & $550$ \\
        STARLINK-1542 & $53.0527$ & $0.0001466$ & $550$ \\
        STARLINK-1553 & $53.0530$ & $0.0001205$ & $550$ \\
        STARLINK-1029 & $53.0533$ & $0.0001596$ & $550$ \\
        STARLINK-1658 & $53.0536$ & $0.0001295$ & $550$ \\
        STARLINK-2039 & $53.0540$ & $0.0001461$ & $550$ \\
        STARLINK-1219 & $53.0543$ & $0.0001173$ & $550$ \\
        STARLINK-2549 & $53.0548$ & $0.0001459$ & $550$ \\
        STARLINK-3548 & $53.2142$ & $0.0000983$ & $550$ \\
        STARLINK-3742 & $53.2157$ & $0.0001568$ & $550$ \\
        STARLINK-3701 & $53.2160$ & $0.0001214$ & $550$ \\
        STARLINK-3542 & $53.2166$ & $0.0000961$ & $550$ \\
        STARLINK-3541 & $53.2170$ & $0.0001236$ & $550$ \\
        STARLINK-3560 & $53.2174$ & $0.0001512$ & $550$ \\
        STARLINK-3633 & $53.2177$ & $0.0001302$ & $550$ \\
        \hline
    \end{tabular}
    \label{tab:TLE_table}
\end{table}
    

There is around $\num{3}$ km error in the estimation of satellite position using the TLE file and SGP4 orbit propagator model \cite{morales2019orbit}.
Moreover, the satellite velocity error of this orbit propagator model is as large as $\num{3}$ m/s.
Most of the position and velocity error of LEO satellites concentrate on tangential direction and along the radial direction, respectively \cite{zhao2023analysis}. 
We add position error of {$\pm$($\num{2}$ to $\num{3}$) km} in the tangential direction and 200 m in the radial direction with the estimated satellite position utilizing the TLE file and SGP4 orbit propagator model. 
Moreover, we add {$\num{2}$ to $\num{3}$ m/s} velocity error in the radial direction and $\num{0.5}$ m/s velocity error in the tangential direction with the estimated velocity.
Using the orbital parameters, the Doppler shift is calculated for both the base receiver and moving UT. 
The tropospheric and ionospheric errors are added with the Doppler shift using \eqref{eq: 5Saastamoinen} and \eqref{eq: 6Klobuchar}, respectively.
{
Doppler shift measurement error is included based on the SNR value of the received signal mapped from \cite{10297313}.}
{Moreover, }Gaussian noise with zero mean and $\num{0.1}$ Hz standard deviation is added as measurement noise.
The Doppler shift is calculated by utilizing the satellite with an elevation angle greater than 15 degrees.
The base receiver transmits its known position and Doppler shift measurements to the UT using any suitable communication technology. The UT utilizes the Doppler shift measurements from common satellites visible to both the base receiver and UT to estimate its position and velocity. 
We consider the differential Doppler positioning method proposed by Neinavaie \textit{et al.} \cite{dif9843493} with our proposed 3DPose framework using the same Doppler shift measurement data.

{
The weight matrix of the WLS solution involves the inverse of measurement noise covariance of the Doppler shift. The noise covariance for Doppler shift measurement can be estimated from the SNR of the received signal \cite{vincent2020doppler}.
In light of this, we use the SNR of the received signal to generate the weight matrix ($\mathbf{W}$).}
The measurement noise of double-difference Doppler shift measurement is the difference of measurement noise between single-difference Doppler shift measurements. Similarly, the measurement noise of a single-difference Doppler shift measurement is the difference of measurement noise between the base receiver and UT. A similar SNR value of the base receiver and UT indicates a similar measurement noise. 
So, we take the difference in SNR value between the base receiver and UT and normalize them to form the diagonal element of the weight matrix. Let $s^\ell_{\sf B}$ and $s^\ell_{\sf UT}$ be the SNR value for the base receiver and UT, respectively. The weight matrix is formed as follows: 
\begin{equation}
    \mathbf{W}={\sf diag}\left(\frac{{\sf max}(\mathbf{s})-\mathbf{s}}{{\sf max}(\mathbf{s})-{\sf min}(\mathbf{s})}\right),
\end{equation}
where $\mathbf{s}=\mathbf{s}_{\sf B}-\mathbf{s}_{\sf UT}$, $\mathbf{s}_{\sf B}=[s^1_{\sf B},s^2_{\sf B},\hdots,s^L_{\sf B}]^T$, and , $\mathbf{s}_{\sf UT}=[s^1_{\sf UT},s^2_{\sf UT},\hdots,s^L_{\sf UT}]^T$.

\subsection{Positioning Error Analysis}
The positioning error comparison of the proposed 3DPose framework with the existing \cite{dif9843493} method is tabulated in Table~\ref{tab: comparison table} regarding RMSE in NEU coordinates for scenario-I, scenario-II, and scenario-III.
From the tabulated data, it is obvious that for scenario-I, the RMSE for {the} existing differential Doppler positioning method \cite{dif9843493} are $\num{0.405}$ m, $\num{1.092}$ m, and $\num{3.066}$ m in {the} north, east, and up {directions}, respectively. 
The RMSE for Vanilla double-difference for scenario-I are $\num{0.396}$ m, $\num{0.940}$ m, and $\num{2.992}$ m in {the} north, east, and up {directions}, respectively. 
In contrast, for our proposed 3DPose framework based on LS reports RMSE for scenario-I as $\num{0.108}$ m, $\num{0.246}$ m, and $\num{0.247}$ m in {the} north, east, and up {directions}, respectively.
For our proposed 3DPose framework based on WLS reports RMSE for scenario-I as $\num{0.108}$ m, $\num{0.245}$ m, and $\num{0.244}$ m in {the} north, east, and up {directions}, respectively.

\begin{table}[t]\renewcommand{\arraystretch}{1.1}
\caption{Comparison of positioning performance in NEU frame}
    \centering
    \begin{tabular}{c c c c c}
    \hlineB{3}
    \rowcolor[HTML]{C0C0C0}
     & \textbf{Positioning} & \multicolumn{3}{c}{\cellcolor[HTML]{C0C0C0} \textbf{RMSE in NEU coordinate (m)}} \\
    \rowcolor[HTML]{C0C0C0}
       \multirow{-2}{*}{\textbf{\cellcolor[HTML]{C0C0C0} Scenario}} &\textbf{\cellcolor[HTML]{C0C0C0} algorithm} &\cellcolor[HTML]{E9EBEF}  \textbf{North(N)} &\cellcolor[HTML]{E9EBEF}  \textbf{East (E)} &\cellcolor[HTML]{E9EBEF} \textbf{Up (U)} \\
         \hline
         \hline
         &Differential Doppler &&& \\
         &positioning method \cite{dif9843493} & \multirow{-2}{*}{$\num{0.405}$} & \multirow{-2}{*}{$\num{1.092}$} & \multirow{-2}{*}{$\num{3.066}$} \\ [2pt]
         &Vanilla double-difference &&& \\
         &based on WLS &\multirow{-2}{*}{$\num{0.396}$} &\multirow{-2}{*}{$\num{0.940}$} &\multirow{-2}{*}{$\num{2.992}$} \\[2pt]
         &3DPose Framework &&& \\
         &based on LS&\multirow{-2}{*}{$\num{0.108}$}& \multirow{-2}{*}{$\num{0.246}$}& \multirow{-2}{*}{$\num{0.247}$} \\  [2pt]   
         &3DPose framework &&& \\    
         \multirow{-7}{*}{Scenario-I}&based on WLS & \multirow{-2}{*}{$\textbf{0.108}$} & \multirow{-2}{*}{$\textbf{0.245}$} & \multirow{-2}{*}{$\textbf{0.244}$} \\
         \hline
         &Differential Doppler &&& \\
         &positioning method \cite{dif9843493} & \multirow{-2}{*}{$\num{2.673}$} & \multirow{-2}{*}{$\num{8.430}$} & \multirow{-2}{*}{$\num{16.759}$} \\[2pt]
         &Vanilla double-difference &&& \\
         &based on WLS &\multirow{-2}{*}{$\num{2.331}$} & \multirow{-2}{*}{$\num{8.084}$}&\multirow{-2}{*}{$\num{16.593}$} \\[2pt]
         &3DPose Framework &&& \\
         &based on LS&\multirow{-2}{*}{$\num{0.137}$}& \multirow{-2}{*}{$\num{0.421}$}& \multirow{-2}{*}{$\num{0.414}$} \\[2pt]
         &3DPose framework &&& \\
         \multirow{-7}{*}{Scenario-II}&based on WLS& \multirow{-2}{*}{$\textbf{0.132}$} & \multirow{-2}{*}{$\textbf{0.356}$} & \multirow{-2}{*}{$\textbf{0.318}$} \\
         \hline
         &Differential Doppler &&& \\
         &positioning method \cite{dif9843493} & \multirow{-2}{*}{$\num{9.890}$} & \multirow{-2}{*}{$\num{38.609}$} & \multirow{-2}{*}{$\num{41.831}$} \\[2pt]
         &Vanilla double-difference &&& \\
         &based on WLS &\multirow{-2}{*}{$\num{9.577}$} &\multirow{-2}{*}{$\num{37.907}$} &\multirow{-2}{*}{$\num{41.206}$} \\[2pt]
         &3DPose framework &&& \\
         &based on LS&\multirow{-2}{*}{$\num{0.739}$} &\multirow{-2}{*}{$\num{1.135}$} &\multirow{-2}{*}{$\num{1.137}$} \\[2pt]
         &3DPose framework &&& \\
         \multirow{-7}{*}{Scenario-III}&based on WLS & \multirow{-2}{*}{$\textbf{0.666}$} & \multirow{-2}{*}{$\textbf{0.599}$} & \multirow{-2}{*}{$\textbf{0.808}$} \\
         \hlineB{3} 
    \end{tabular}
    \label{tab: comparison table}
\end{table}

\begin{figure}[t]
    \centering
    \includegraphics[width=1\linewidth]{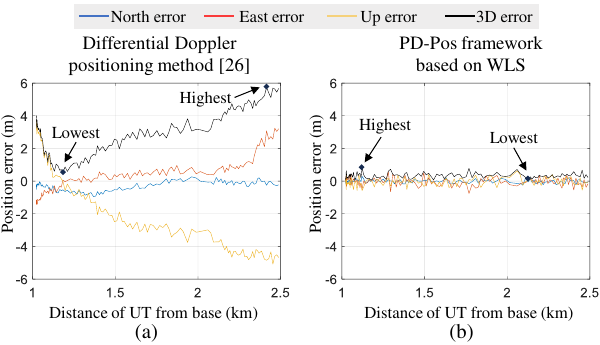}
    \caption{{Time series positioning error for scenario-I. (a) Differential Doppler positioning method [26]. (b) Proposed 3DPose framework based on WLS.}}
    \label{fig: NED3D_1}
\end{figure}

For scenario-II, {the} existing differential Doppler positioning method \cite{dif9843493} {achieves} RMSE of $\num{2.673}$ m, $\num{8.430}$ m, and $\num{16.759}$ m in {the} north, east, and up {directions}, respectively. 
The RMSE for Vanilla double-difference for scenario-II are $\num{2.331}$ m, $\num{8.084}$ m, and $\num{16.593}$ m in {the} north, east, and up {directions}, respectively.
Although the distance between the base receiver and UT increases, our proposed 3DPose framework effectively reduces the position error. The proposed 3DPose framework based on LS obtains RMSE of $\num{0.137}$ m, $\num{0.421}$ m, and $\num{0.414}$ m in the north, east, and up direction, respectively, for \mbox{scenario-II}. The proposed 3DPose framework based on WLS obtains RMSE of $\num{0.132}$ m, $\num{0.356}$ m, and $\num{0.318}$ m in the north, east, and up direction, respectively, for scenario-II. 

With a long baseline, the existing differential Doppler positioning method \cite{dif9843493} can not mitigate the ephemeris error effectively, which causes a large positioning error. The positioning errors for scenario-III of {the} existing differential Doppler positioning method \cite{dif9843493} are $\num{9.890}$ m, $\num{38.609}$ m, and $\num{41.831}$ m in {the} north, east, and up {directions}, respectively. 
The RMSE for Vanilla double-difference for scenario-III are $\num{9.577}$ m, $\num{37.907}$ m, and $\num{41.206}$ m in {the} north, east, and up {directions}, respectively.
In contrast, our proposed 3DPose framework performs better positioning accuracy for \mbox{scenario-III}. The RMSE for scenario-III of our proposed 3DPose framework based on LS is $\num{0.739}$ m, $\num{1.135}$ m, and $\num{0.137}$ m in {the} north, east, and up {directions}, respectively. The RMSE for \mbox{scenario-III} of our proposed 3DPose framework based on WLS are $\num{0.666}$ m, $\num{0.599}$ m, and $\num{0.808}$ m in {the} north, east, and up {directions}, respectively.

Compared to all the experimental results of different methods, the proposed 3DPose based on WLS outperforms others for all three scenarios.
These results prove that the proposed ephemeris error correction algorithm (Algorithm~\ref{alg: error}) can adequately estimate the ephemeris error and, at the same time, can minimize it accordingly. 
Therefore, we can claim that the Doppler shift measurements are free from ephemeris error.
Overall, the proposed framework can estimate the precise and robust position of the UT considering the corrected Doppler shift measurement.

\subsection{Time Series Positioning Error Analysis}
We have performed a time series positioning error analysis of our proposed 3DPose framework based on WLS with the existing \cite{dif9843493} method to get a better intuition of the performance between different methods regarding NEU coordinates and {3-D} {coordinates}.
Fig.~\ref{fig: NED3D_1} shows the NEU and 3-D time series positioning error of the trajectory for scenario-I. The maximum {3-D} position error does not exceed more than $\num{1}$ m for our proposed 3DPose framework based on WLS. Whereas, {3-D} position error for {the} existing differential Doppler positioning method \cite{dif9843493} is as large as $\num{6}$ m.  
\begin{figure}[t]
    \centering
    \includegraphics[width=1\linewidth]{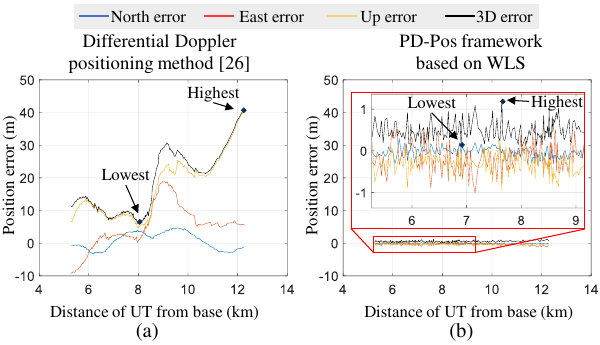}
    \caption{{Time series positioning error for scenario-II. (a) Differential Doppler positioning method [26]. (b) Proposed 3DPose framework based on WLS.}}
    \label{fig: NED3D_2}
\end{figure}
\begin{figure}[t]
    \centering
    \includegraphics[width=1\linewidth]{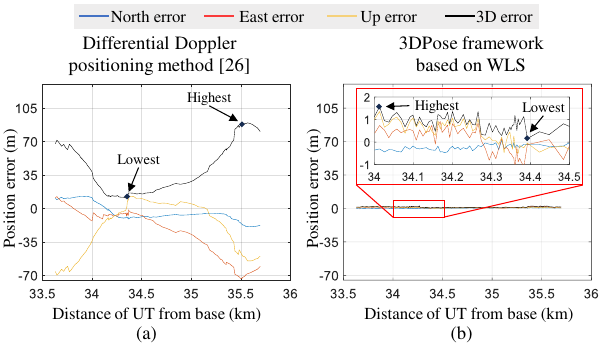}
    \caption{{Time series positioning error for scenario-III. (a) Differential Doppler positioning method [26]. (b) Proposed 3DPose framework based on WLS.}}
    \label{fig: NED3D_3}
\end{figure}The ephemeris error for the base receiver and UT is the same for a relatively short distance between them. Hence, the conventional positioning algorithm's positioning error is low. However, the positioning error increases as the distance between the base receiver and UT increases.

The NEU and {3-D} position errors {for scenario-II} are shown in Fig.~\ref{fig: NED3D_2}. Whereas, the minimum {3-D} position error for the existing differential Doppler positioning method \cite{dif9843493} is $\num{6}$ m, while {3-D} position error for our proposed 3DPose framework based on WLS does not exceed $\num{1.25}$ m. The maximum position error for scenario-II is $\num{41}$ m for the existing differential Doppler positioning method \cite{dif9843493}.

\begin{figure}[t]
    \centering
    \includegraphics[width=0.9\linewidth]{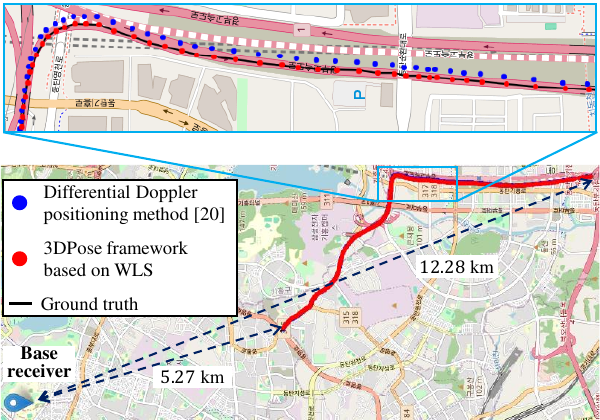}
    \caption{Evaluation map for scenario-II.}
    \label{fig:map_2}
\end{figure}
\begin{figure}[t]
    \centering
    \includegraphics[width=0.9\linewidth]{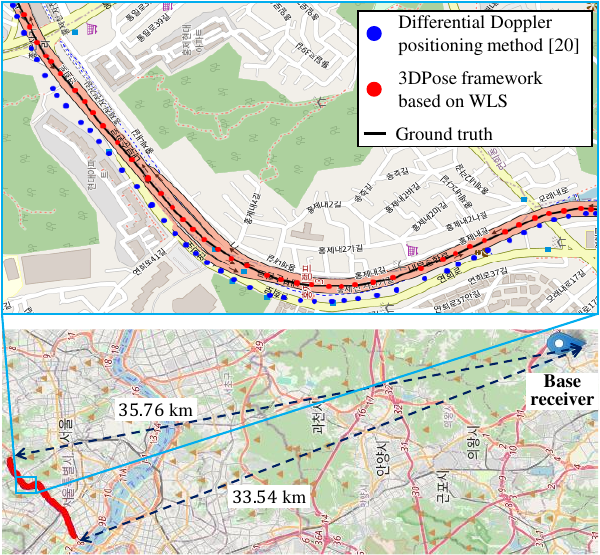}
    \caption{Evaluation map for scenario-III.}
    \label{fig:map_3}
\end{figure}

{
For a long baseline, the existing differential Doppler positioning method \cite{dif9843493} suffers significant position error due to the lack of handling uncertainty in Doppler shift measurement error. The NEU and {3-D} position error for the trajectory of long baseline scenario-III is shown in Fig.~\ref{fig: NED3D_3}. From the figure, it is obvious that there is uncertainty in positioning error with the changes of baseline, and the position error of the existing differential Doppler positioning method [20] for scenario-III varies from 90 m to 10 m. To mitigate this uncertainty in positioning error, our proposed ephemeris error correction algorithm recalculates the Doppler shift error due to the ephemeris error for UT at the initial estimated position. The figure shows that most of the time, the 3-D position error for our proposed 3DPose framework based on WLS is between 0.16 m and 1.78 m. Therefore, we can claim that our proposed 3DPose positioning framework is robust in terms of handling uncertainty in Doppler shift measurement error.}

\subsection{Evaluation Map}

The position outcomes of {the} existing differential Doppler positioning method \cite{dif9843493} and our proposed 3DPose framework based on WLS are shown in Figs.~\ref{fig:map_2} and \ref{fig:map_3} for \mbox{scenario-II} and scenario-III, respectively. The ground truth and the outcome of the existing differential Doppler positioning method \cite{dif9843493} and our proposed 3DPose framework based on WLS are marked in black, blue, and red color, respectively.
From the evaluation map, it is shown that the position outcome of the proposed 3DPose framework based on WLS seems far closer to the ground truth than the existing differential Doppler positioning \cite{dif9843493}.

From all the results in the case of all three scenarios, it is obvious that the proposed 3DPose framework improves UT's positioning accuracy remarkably, even in a situation where the baseline is large. 
{The significant positioning accuracy of UT is due to the correct estimation of the ephemeris error by the proposed ephemeris error correction algorithm.
This algorithm recalculates the ephemeris error induced in the Doppler shift measurements at the estimated UT position. Utilizing these corrected Doppler shift measurements, a precise UT position can be estimated by the WLS solution.}
{Although the numerical results depend on the evaluation setting, there is no significant difference in position outcomes in the case of all three scenarios.}
Thus, we can claim that our proposed 3DPose framework effectively minimizes all the error terms (ionospheric, tropospheric, clock synchronization, and ephemeris) that ensure better positioning accuracy in both short and long baselines.
Therefore, the proposed 3DPose framework outperforms the existing differential Doppler positioning method \cite{dif9843493} in terms of positioning accuracy.

\section{Conclusion and Future Directions}
\label{sec: Conclusion}
In this paper, we have proposed the 3DPose framework to address the positioning error caused by the clock synchronization deficiency between the base receiver and UT as well as the long baseline.
Firstly, the calculation of double-difference Doppler shift measurement has been done using the single-difference Doppler shift measurement with the measurement from a reference satellite to eliminate UT clock drift and to reduce other error terms.
Subsequently, the Doppler shift measurement error due to the ephemeris error has been recalculated using the developed ephemeris error correction algorithm to handle the positioning error of UT with a long baseline. 
A performance comparison has been done with the existing differential Doppler positioning method and the different variants of the proposed 3DPose framework.
The experimental results prove that the proposed 3DPose framework based on WLS outperforms the existing differential Doppler positioning method.

{Most of the position errors of the LEO satellite mainly condense in the along-track direction. The proposed 3DPose framework models the Doppler shift measurement error for UT considering the LEO satellite position error in the along-track direction and improves the positioning accuracy of UT. However, there is a small portion of the satellite's position error along the cross-track and radial directions that may influence position accuracy.}
%
%
{Future work will involve the analysis of the impact of the real-world applicability of the proposed 3DPose framework considering satellite position error along the cross-track and radial directions as well as the development of satellite selection algorithms to find the satellite with good geometry to ensure better coverage as well as UT positioning.}

\bibliographystyle{IEEEtran}
\bibliography{ref}

\end{document}